\documentstyle[preprint,tighten,aps,axodraw,eqsecnum,psfig,epsf,prd]{revtex}
\begin{document}
\preprint{\baselineskip 18pt{\vbox{\hbox{CERN-TH/99-123}
\thispagestyle{empty}
\hbox{NTUA--73/99}\hbox{OUTP-98-89P}\hbox{hep-ph/9905272}}}}
\title{On magnetic catalysis in even-flavor QED${}_3$}
\vspace{15mm}
 \author{K. Farakos, G. Koutsoumbas}
\address{Department of Physics, National Technical University of
Athens,\\ 
Zografou Campus, 157 80 Athens, GREECE}

\author{N.E. Mavromatos} 
\vspace{15mm}
\address{CERN, Theory Division, Geneva 23 CH-1211, Switzerland,\\
and Dept. of Physics, Theoretical Physics, 
Univ. of Oxford, 1 Keble Rd., OX1 3NP, U.K.} 

\vspace{1mm}

\author{Arshad Momen}
\vspace{15mm}
\address{Dept. of Physics, Theoretical Physics, 
Univ. of Oxford,\\ 1 Keble Rd., OX1 3NP, U.K. }
\vspace{1mm}

\maketitle

\begin{abstract}
In this paper, we discuss the role of an external magnetic field on
the dynamically generated fermion mass in even-flavor 
$QED$ in three space-time dimensions. 
Based on some reasonable approximations, 
we present analytic arguments 
on the fact that, for weak fields, the magnetically-induced 
mass increases quadratically
with increasing field, 
while at strong fields one crosses over to a mass
scaling logarithmically with the external field. We also confirm this
type of scaling behavior through quenched lattice calculations using the
non-compact version for the gauge field. Both the zero and finite temperature
cases are examined. 
A preliminary study of the fermion condensate in the presence of magnetic
flux tubes on the lattice is also included.   
\end{abstract}
\vspace{5mm}

\newcommand{\be}{\begin{equation}}
\newcommand{\ee}{\end{equation}}
\newcommand{\bea}{\begin{eqnarray}}
\newcommand{\eea}{\end{eqnarray}}
\newcommand{\real}{{\rm l}\! {\rm R}}
\newcommand{\ra}{\rightarrow}
\newcommand{\tr}{\mbox{ tr }}
\newcommand{\Tr}{\mbox{ Tr }}
\newcommand{\al}{\alpha}
\newcommand{\bt}{\beta}
\newcommand{\bz}{{\bar{z}}}

\newcommand{\del}{\Delta}
\newcommand{\Th}{\Theta}
\newcommand{\td}{\tilde{\del}}
\newcommand{\g}{\gamma}
\newcommand{\m}{\mu}
\newcommand{\n}{\nu}
\newcommand{\bchi}{\bbox{\chi}}
\newcommand{\nn}{\nonumber}

\def\lsim{\mathrel{\rlap {\raise.5ex\hbox{$ < $}}
{\lower.5ex\hbox{$\sim$}}}}

\section{Introduction}

The particle mass generation  via dynamical symmetry 
breaking has been a
much-studied scenario in particle physics as well as in 
condensed-matter systems.
In the recent years this phenomenon has been studied in the presence
of background fields, such as constant external magnetic 
fields~\cite{gusynin}-\cite{new}, following and extending
the formalism developed by Schwinger~\cite{schwinger}. 
The formalism has been applied to models that had 
gauge and/or four-fermion interactions. 
It was 
found that such constant background configurations
can enhance the dynamical symmetry breaking by driving the critical
coupling to a smaller value 
and thus catalyzing the symmetry breaking.  
A concrete example of this phenomenon, of relevance to us in this work, 
is the dynamical chiral symmetry 
breaking of chiral symmetry in  
massless QED ( in three and four dimensions) in the presence 
of an external 
magnetic field~\cite{gusynin,leung,shpagin,kostas,lattice} 
where 
the dynamically generated fermion mass depends 
on the value of the external field. 

The magnetically catalyzed mass generation 
in (2+1) dimensional QED 
may have interesting condensed-matter applications~\cite{kostas,lattice},
given the suggestions 
that 
high-temperature superconductors can be described effectively by 
field theories like 
QED$_3$ \cite{NN} or by
non-Abelian gauge models based on the group
$SU(2) \times U(1)$ \cite{fm,kostas}
\footnote{The relativistic
(Dirac) nature of the fermion fields is justified by the fact that
they  describe the excitations about the {\em
nodes} of a $d$-wave superconducting gap.}.
Indeed, there is
experimental evidence 
for the opening of a second (superconducting) gap at the nodes  of the
gap in certain d-wave superconductors in the presence of strong
external magnetic fields \cite{krishna}.
As remarked in \cite{lattice2},
in the context of condensed-matter-inspired models, 
the scaling of the thermal conductivity 
with the external field is different between the gauge~\cite{kostas,lattice} 
and four-fermion 
theories~\cite{ssw}. Thus, a detailed study of the 
magnetically-induced chiral symmetry breaking phenomenon
in the context of $QED_3$ is phenomenologically desirable,
given that such studies
may lead to 
more detailed experiments in the spirit of \cite{krishna}, 
that can probe deep in the structure of the novel high-temperature
superconductors.
 
In (2+1) dimensions chiral symmetry can be defined only if the number of 
fermion flavors is even \cite{appelquist}. This fact is relevant for  
a planar high-$T_c$ superconducting antiferromagnetic 
system~\cite{NN,aitch} which comprises of 
two sublattices.  Within 
a generalized~\cite{fm} spin-charge separation framework~\cite{Anderson}, 
there will be  
two species of charged fermion excitations (called holons), one
associated with each sublattice~\cite{NN,fm}.  Finally,
the  (2+1) dimensional theory with
even number of fermion flavors~\cite{kostas} can be viewed as a
dimensional reduction  of the four-dimensional effective
Lagrangian  of \cite{hong}.

In $QED_3$, the magnetic catalysis of the chiral symmetry breaking  
for strong external fields
is  established  by looking at the Schwinger-Dyson equations 
\cite{shpagin,kostas}. In these works the Landau level formalism was used 
to truncate the fermion propagators to the lowest Landau level.
This formalism is satisfactory for certain aspects of the 
magnetic catalysis for strong magnetic fields~\cite{kostas}, 
but for weak fields the result can definitely be questioned, given that 
in that case the spacing between Landau levels becomes small, and
one effectively deviates from the lowest Landau level description. 
Recently two of us~\cite{us} have looked at the r\^ole of higher Landau levels and showed that they contribute by inducing a 
(parity-violating) magnetic moment which scales 
with the applied magnetic field. 
Moreover the r\^ole of higher Landau levels in inducing a 
critical temperature even in the free fermion case,
under certain circumstances, 
was emphasized in \cite{lattice}. 
For all of the above reasons it is important to incorporate 
the effects of {\it all} 
the higher Landau levels in the Schwinger-Dyson formalism, 
avoiding the use of the mean field Landau level decomposition 
altogether. This is what we shall attempt to do in the
first (analytic) part of this paper. 
We shall compare these results by performing some 
preliminary (quenched) lattice analyses
in the second part of the paper. In the latter respect we have to 
mention that the quenched approximation for fermions employed
here allows for the ladder gauge quantum fluctuations in the fermion 
free energy to be incorporated, but prevents the use of internal 
fermion loops, as the treatment of the latter requires 
an algorithm for treating dynamical fermions which  
is currently under construction.

The paper is organized as follows. In section 2 we give a brief
review  of the $SU(2) \times U_S(1)$ model of \cite{fm}, as
well as the Dirac algebra in three-dimensional spacetime with an 
even number of fermion flavors. In section 3 we review the
Schwinger-Dyson (SD) equation for the fermion propagator in the
absence of the
the external magnetic field. In section 4 we present the results
for the case of strong external fields, where a  
logarithmic scaling of the induced condensate with the 
external magnetic field occurs. In section 5 we present the 
SD equations for the weak magnetic fields ignoring the photon
polarization to make contact with the lattice result presented in the
second half of the paper. We show that under certain approximations,
the scaling behaviour of the condensate with the
external magnetic field can be found. In the next section, 
we attempt to go beyond the quenched approximation 
analytically,
by including the photon polarization and modify 
accordingly the Schwinger-Dyson equations.
The analysis becomes very complicated to be handled analytically
for finite temperatures, and this is the reason why we 
turn to the lattice formulation of the problem 
in section 7, where we set up the formalism and 
relevant notations. 
In section 8 the lattice results are presented 
for both zero and finite 
temperatures; in addition, a preliminary extension of the results to 
the non-uniform magnetic field cases is attempted by examining 
the magnetic catalysis phenomenon in the case of flux tubes. 
This model may constitute a prototype for the study of the effects 
of electromagnetic vortices in condensed matter systems, which are 
of relevance to high-temperature superconductivity.
Conclusions and outlook are presented in section 9. 

\section{The model and its symmetries}

The $SU(2) \times U(1)$ model of \cite{farak} is a toy model for
dynamical electroweak gauge symmetry breaking in three dimensions,
while in the context of condensed-matter systems, the $SU(2) \times
U_S(1)$ model  of \cite{fm} is based on a gauged 
{\it particle-hole symmetry}, via a suitable extension of the  
spin-charge separation~\cite{Anderson}. The 
holons transform as a doublet under the $SU(2)$ 
(particle-hole) symmetry. In this respect the model is different 
from other $SU(2) \times U(1)$ spin-charge separated theories,
which are based on either direct gauging of genuine spin rotation $SU(2)$ 
symmetries~\cite{leewen}, or non-Abelian bosonization 
techniques~\cite{marchetti,yulu}. The phase diagram  
of the model of \cite{fm}, and the associated symmetry-breaking patterns, 
are quite different from these other models.

The three-dimensional 
continuum Lagrangian of the model is given ( in Euclidean
metric, which we use hereafter) by ~\cite{farak,kostas}, 
\be 
{\cal L} = -\frac{1}{4}(F_{\mu\nu})^2 -\frac{1}{4}({\cal G}_{\mu\nu})^2
+ {\overline \Psi}_i D_\mu\gamma _\mu \Psi_i -m {\overline \Psi}_i \Psi_i 
\label{contmodel}
\ee
where $D_\mu = \partial _\mu -ig_1 a_\mu^S - ig_2 \sigma^a B_{a,\mu} $,
and $F_{\mu\nu}$, ${\cal G}_{\mu\nu}$ are the corresponding field
strengths
for an abelian (``statistical")  $U_S(1)$ gauge field $a^S_\mu$ and 
a non-abelian (``spin") SU(2) gauge field $B^a_\mu$, respectively. 
Due to the antiferromagnetic nature of the condensed matter system
the  
fermions $\Psi_i$ are four-component spinors, $ i = 1, \cdots N$. 
We note that $\Psi_i$ may be written as 
\be
\Psi_i \equiv \left( \begin{array}{c} \Psi_{i1} \\ 
\Psi_{i2}\end{array}\right).
\ee
Then the Lagrangian decomposes into two parts,
one for $\Psi_{i1}$ and one for $\Psi_{i2},$ 
which will be called ``fermion species" in the sequel.
The presence of the even
number of fermion species allows us to define chiral symmetry and
parity in three dimensions\cite{appelquist}, which we discuss below.
The bare mass $m $ term is parity conserving and  
has been added by hand 
in the Lagrangian (\ref{contmodel}). In the model of \cite{fm,kostas},
this term is  generated dynamically  via 
the formation of the fermion condensate
$<{\overline \Psi} \Psi >$ by the strong 
$U_S(1)$ coupling. However, for our purposes, the details of the 
dynamical mass generation is not important and hence  
it will be sufficient to include a bare mass term for the 
holons  representing the  
mass generated by the (strongly coupled) $U_S(1)$ interactions
in the superconducting phase. 

In what follows we shall ignore for simplicity the non abelian 
gauge group structure and concentrate only in the Abelian 
model in the presence of an {\it external } electromagnetic 
field, which should not be confused with the 
statistical abelian gauge field $U_S(1)$. 
The incorporation of the gauged $SU(2)$ structure 
leads to a much richer phase structure~\cite{mcneill,us}
and we reserve the discussion for future publication.

For even-flavour models a convenient representation for 
the $\gamma _\mu $, $\mu =0,1,2$,  
matrices 
is the reducible $4 \times 4$ representation of the 
Dirac algebra in three dimensions~\cite{appelquist}:
\bea
&~&\gamma ^0=\left(\begin{array}{cc} 
i{\bf \sigma}_3 \qquad {\bf 0} \\
{\bf 0} \qquad -i{\bf \sigma}_3 
\end{array} \right) 
\qquad \gamma ^1 = \left(\begin{array}{cc} 
i{\bf \sigma}_1 \qquad {\bf 0} \\
{\bf 0} \qquad -i{\bf \sigma}_1 \end{array} \right) \nn \\
&~&\gamma ^2 = \left(\begin{array}{cc} i{\bf \sigma}_2 
\qquad {\bf 0} \\{\bf 0} \qquad -i{\bf \sigma}_2 \end{array} \right) 
\label{reduciblerep}
\eea
where 
${\bf \sigma}$ are $2 \times 2$ Pauli matrices
and the (continuum) space-time is taken to have Euclidean signature.

As well known~\cite{appelquist} there exist two $4 \times 4 $ matrices 
which anticommute with $\gamma _\mu$, $\mu=0,1,2$: 
\be
\gamma _3 =\left(\begin{array}{ll} 0 \qquad {\bf 1} \\
{\bf 1} \qquad 0 \end{array}\right), \qquad 
\gamma _5 =i\left(\begin{array}{ll} 0 \qquad {\bf 1} \\
{\bf -1} \qquad 0 \end{array}\right)
\label{gammamatr}
\ee
where the substructures are $2 \times 2$ matrices.
These are the generators of the `chiral' symmetry for 
the massless-fermion theory: 
\bea
&~&   \Psi \rightarrow exp(i\theta \gamma _3) \Psi \nn \\
&~&   \Psi \rightarrow exp(i\omega \gamma _5) \Psi. 
\label{chiral}
\eea
Note that these 
transformations do not exist in the fundamental two-component 
representation
of the three-dimensional Dirac algebra, and therefore 
the above symmetry is valid for theories 
with even fermion species only. 

For later use we note that
the Dirac algebra in $(2 +1)$ dimensions satisfy the identity:
\bea 
&~&\gamma ^\mu \gamma ^\nu = -\delta ^{\mu\nu} 
- \tau _3 \epsilon ^{\mu\nu\lambda} \gamma ^\lambda \quad ; \quad  
\tau _3 \equiv i\gamma_3\gamma _5= \left(\begin{array}{cc} {\bf 1}  
\qquad {\bf 0} \\{\bf 0} \qquad {\bf -1}  \end{array} \right) \nn \\
&~& \gamma ^\mu \gamma ^\lambda \gamma ^\mu = \gamma ^\lambda \nn \\
&~&\gamma ^\mu \gamma ^0 \gamma ^i \gamma ^j \gamma ^\mu = - \delta ^{ij}\gamma ^0 - 3 \tau _3 \epsilon ^{ij} \nn \\
&~& \gamma ^\mu \gamma ^i \gamma ^j \gamma ^\mu = - 3\delta ^{ij} - 
\tau _3 \gamma ^0 \epsilon ^{ij} \nn \\ 
&~&\gamma ^\mu \gamma ^j \gamma ^i \gamma ^k \gamma ^\mu 
= - \delta ^{ij}\gamma ^k - \delta ^{ik}\gamma ^j + \delta ^{jk}\gamma ^i
\label{identities}
\eea
which is specific to three dimensions only.
Here the Greek indices
are space time indices, and 
repeated indices denote summation.

{\it Parity} in this formalism is defined as the transformation:
\be
P:~\Psi (x^0, x^1, x^2) \rightarrow -i\gamma^3 \gamma ^1 \Psi (x^0, -x^1, x^2) \label{parity}
\ee 
and it is easy to see that a parity-invariant mass term for $\Psi$ 
amounts to masses with {\it opposite} signs between the two 
species~\cite{appelquist}, while a parity-violating one 
corresponds to masses of equal signs. 

The set of generators 
\be
{\cal G} = \{ {\bf 1}, \gamma _3, \gamma _5, 
\Delta \equiv i\gamma_3\gamma _5 \}
\label{generators}
\ee
form~\cite{farak,fm} 
a global $U(2) \simeq SU(2) \times U_S(1)$ symmetry. 
The identity matrix ${\bf 1}$ generates the $U_S(1)$ subgroup, 
while 
the other three form the SU(2) part of the group. 
The currents corresponding to the above transformations 
are:
\be
   J_\mu^\Gamma = {\overline \Psi} \gamma _\mu \Gamma \Psi 
\qquad \Gamma =\gamma _3,\gamma _5, i\gamma _3\gamma _5 
\label{currents}
\ee
and are {\it conserved} in the {\it absence} of a fermionic {\it mass}
term. 
It can be readily verified that the corresponding charges
$Q_\Gamma \equiv \int d^2x \Psi ^\dagger \Gamma \Psi $ lead
to an $SU(2)$ algebra~\cite{farak}:
\bea
 &~& [Q_3, Q_5]=2iQ_\Delta \qquad [Q_5,Q_\Delta ]=2iQ_3 \nn \\
&~&  [Q_\Delta, Q_3]=2iQ_5 
\label{chargealgebra}
\eea
In the presence of a mass term, these currents are not conserved:
\be
       \partial ^\mu J_\mu^\Gamma = 2m {\overline \Psi } \Gamma \Psi,
\label{anomaly}
\ee
while the current corresponding to the generator ${\bf 1}$ 
is {\it always } conserved, even in the presence of a fermion 
mass. The situation is parallel to the treatment of the $SU(2) \times  SU(2)$
chiral symmetry breaking in low-energy QCD and the partial conservation of
axial current ( PCAC).
The bilinears
\bea
&~&{\cal A}_1 \equiv {\overline \Psi}\gamma _3 \Psi,  
\qquad {\cal A}_2 \equiv {\overline \Psi}\gamma _5 \Psi,  
\qquad {\cal A}_3 \equiv {\overline \Psi}\Psi 
\nn \\
&~&B_{1\mu} \equiv {\overline \Psi}\gamma _\mu \gamma _3 \Psi,~
B_{2\mu} \equiv {\overline \Psi}\gamma _\mu \gamma _5 \Psi,~
B_{3\mu} \equiv {\overline \Psi}\gamma _\mu \Delta \Psi,~\mu=0,1,2      
\label{triplets}
\eea
transform as {\it triplets} under $SU(2)$. 
The $SU(2)$ singlets are 
\be 
{\cal A}_4 \equiv {\overline \Psi}\Delta \Psi, \qquad 
B_{4,\mu} \equiv {\overline \Psi}\gamma _\mu \Psi 
\label{singlets}
\ee
i.e. the singlets are the parity violating mass term, 
and the four-component fermion number. 

We now notice that in the case 
where the fermion condensate ${\cal A}_3$ is generated 
dynamically, energetics 
prohibits the generation of a parity-violating 
gauge invariant $SU(2)$ term~\cite{vafa}, and so 
a parity-conserving mass term necessarily breaks~\cite{kostas}
the $SU(2)$ group down to a $\tau_3$-$U(1)$ sector~\cite{NN}, generated
by the $\sigma_3$ Pauli matrix in two-component notation. 
 Upon coupling the system to external electromagnetic 
potentials, this phase with massive fermions shows  
{\it superconductivity}. 
The superconductivity is strongly type II~\cite{NN,kostas} as 
the Meissner penetration depth of external magnetic 
fields turn out to be  very large,\footnote{ The high-temperature 
superconducting oxides are  strongly type II superconductors.} and hence
the study of the response of the system to 
the external electromagnetic fields is justified.

\section{ The Schwinger-Dyson equation for the Fermion}

$QED_3$ is a superrenormalizable theory which is confining in the
infrared regime. Accordingly, it acts as a simple prototype for the
analysis of 
the chiral symmetry breaking in QCD. The standard tool for
investigating the chiral symmetry breaking are the celebrated
Schwinger-Dyson equations. 
In this section, let us set up the Schwinger-Dyson equations for the 
fermion propagator. 

The Schwinger-Dyson equation concerning the fermion propagator $S_F(p)$ 
(for zero bare fermion mass) is given by:
\be
S_F^{-1}(p) = \gamma \cdot p  - g \int \frac{d^3k}{ ( 2\pi)^3} \gamma^\mu S_F (k)
\Gamma^\nu ( k, p-k) D_{\m \n}(p-k)
\label{2.1}
\ee
where $\Gamma^\n$ is the fermion-photon vertex function and $D_{\m
\n}$ is the exact photon propagator. However, to this order, let us
make the following approximations:

\begin{enumerate} 

\item  Use the bare vertex function, namely
\be
 \Gamma^\nu ( k, p-k) = g \gamma^\n,
\label{2.2}
\ee
so that the gap equation reads:
\be
S_F^{-1}(p) = \gamma \cdot p  - g^2 \int 
\frac{d^3k}{ ( 2\pi)^3} \gamma^\mu S_F (k)
\g^\nu ( k, p-k) D_{\m \n}(p-k)
\label{2.1.5}
\ee

\item

Now, we choose the following ansatz for the {\em full} fermion propagator:
\be
S_F^{-1} (p) = A(p) \g^0 p_0 + B(p) \bbox{ \g \cdot p} + \Sigma(p)
\label{2.3}
\ee
Using this ansatz, let us now perform a trace over the gamma matrices 
in (\ref{2.1}). This gives us the following gap equation
\be
\Sigma(p) =  g^2 \int \frac{d^3 k}{(2\pi)^3} \frac{ \Sigma(k)}{ A^2
k_0^2 + B^2 \bbox{ k^2} + \Sigma^2(k) }\sum_{\m} D_{\m \m} ( p-k)
\label{2.4}
\ee

\item To further simplify the gap equation let us use the zeroth order 
result for the wavefunction renormalization, namely $A(p)= B(p)=1$, 
which is often justified in the large $N$ argument
\cite{appelquist} (see however \cite{ian}) 
so that 
eq.(\ref{2.4}) reads:
\be
\Sigma(p) = g^2 \int \frac{d^3 k}{(2\pi)^3} \frac{ \Sigma(k)}{
(k_0^2 +  \bbox{ k^2} + \Sigma^2(k) )} D_{\m \m} ( p-k)
\label{2.5}
\ee

\item The photon propagator $D_{\m \n}(k)$ can be replaced by the ladder resummed propagator which can be justified in the large-$N$ limit. The resummed 
propagator (in the absence of the magnetic field) is given by 
\be
D_{\m \n}(p) 
= \frac{(\delta_{\m \n} - \frac{ p_\m p_\n}{p^2})}{p^2(1-\Pi(p))}
= \frac{(\delta_{\m \n} - \frac{ p_\m p_\n}{p^2})}{p^2(1+\frac{g^2}{8 p})}
\label{2.5.5}
\ee
\end{enumerate}

The gap equation thus obtained in the absence of the magnetic field 
can be solved using the bifurcation method \cite{appelquist}. There are 
two solutions namely, 
\be
\Sigma_1(p) \sim p^{-8/\pi^2 N}, \qquad \Sigma_2(p) \sim p^{1-8/\pi^2 N}
\label{2.5.6}
\ee
where $N$ is the (large) number of fermion flavours.  
However, it is natural to expect that these solutions will change 
in the presence of the external magnetic field; 
we will discuss this generalization in the following sections.

\section{The Dynamically generated Fermion Mass at Strong Magnetic Fields}

As mentioned above, hereafter we consider only the abelian
gauge group $U_S(1)$ in the presence of an external 
electromagnetic potential $A_\mu^{ext}$, corresponding to 
a constant magnetic field ${\bf B}$, perpendicular to the spatial 
plane. The dynamics is described by the Lagrangian:
\be 
{\cal L} = -\frac{1}{4}(F_{\mu\nu})^2 
+ {\overline \Psi} D_\mu\gamma _\mu \Psi -m {\overline \Psi} \Psi 
\label{contmodel2}
\ee
where $D_\mu = \partial _\mu -ig a_\mu^S - ie A^{ext}_\mu $.
The mass $m$ here should be viewed as an (infrared ) 
regulator mass. In the 
dynamical mass generation scenario investigated below via the SD method
$m$ should be set to zero, given that the dynamics of  
the gauge field and the 
magnetic field are both responsible for the 
appearance of a mass in the fermion propagator.
For the lattice analysis, on the other hand,
the presence of an initial small `bare' regulating mass 
$m \ne 0$ appears 
necessary~\cite{lattice}.

We commence our analysis by noting that 
the presence of an external magnetic field, perpendicular 
to the spatial plane $x_1x_2,$ breaks 
Lorentz and translational invariance. 
The configuration space form of the 
fermion two-point function 
$G(x,y)$ for the three-dimensional problem at 
hand has the generic form~\cite{shpagin}:
\begin{equation} 
G(x,y) = 
{\rm exp}\left(\frac{ie}{2}(x-y)^\mu A_\mu^{{\rm ext}}(x+y)\right){\tilde G}(x-y)
\label{confprop}
\end{equation}
where $A_\mu ^{{\rm ext}}$ denotes 
the external electromagnetic potential, corresponding to 
a constant homogeneous magnetic field perpendicular to the spatial plane 
$x_1x_2:$
$A_\mu^{{\rm ext}} =\left(0, -\frac{B}{2}x_2, \frac{B}{2}x_1 \right)$ (in an obvious notation). 
The field-dependent phase factor in (\ref{confprop}) breaks
translational invariance, implying that, 
in general, $G(x,y)$ does not admit a Fourier transform 
expressible in terms of a single momentum (vector) variable $\bbox{k}.$
 
The translationally-invariant part ${\tilde G}(x-y)$ has 
a Fourier transform 
$\tilde{S}_F(k)$ of the form~\cite{schwinger}:
\be
\tilde{S}_F(k) =i \int^\infty_0 ds e^{- s ( k_0^2 + m^2 +\bbox{k}^2 \frac{\tanh
z}{z})} [ ( m- \g \cdot  k) - i ( \g_1 k_2 - \g_2 k_1) \tanh z ]
( 1- i \g_1 \g_2 \tanh z)
\label{schwing}
\ee
where $m$ is the mass of the fermion,
and $z= s~eB$.  Note that we are distinguishing
between the coupling constant $g$ for the statistical $U(1)$ gauge
field and the electromagnetic charge $e$.
The Schwinger propagator admits the
following expansion in terms of the Landau levels~\cite{chodos}:
\be
\tilde{S}_F(k) \equiv i e^{-\frac{{\bbox{ k}}^2}{eB} }\sum^\infty_{n=0}
\frac{(-1)^n D_n (k_0,{\bbox{k}})}{k_0^2 +m^2 + 2 e n B},~~~B \ne 0,
\label{1.2}
\ee
with
\bea
&&D_n(k_0, {\bf k}) \equiv \nn \\
&& ( m - \g_0 k_0) \left[ ( 1- i \g_1 \g_2) 
L_n(\frac{2 {\bbox{ k}}^2}{eB}) - ( 1+ i \g_1 \g_2) L_{n-1}(\frac{2{\bbox{
k}}^2} {eB})
\right] + 4 ( \bbox{\g \cdot k}) L_{n-1}^1 (\frac{2 {\bbox{ k}}^2}{eB})
\label{1.3bf}\\
&=& ( m- \g_0 k_0) \left[ L_n^{-1}( \frac{2\bbox{k}^2}{eB}) + i \tau_3 \g_0
\left( L_n( \frac{2\bbox{k}^2}{eB} ) + L_{n-1}(\frac{2 \bbox{k}^2}{eB} ) 
\right) \right] + 4
( \bbox{\g \cdot k}) L_{n-1}^1(\frac{2\bbox{k}^2}{eB})
\label{1.3}
\eea

For $QED_3$ the scaling of the dynamically generated fermion mass with
the external magnetic field had been discussed by Shpagin
\cite{shpagin} and two of us \cite{kostas}.
Let us begin with the case when the external magnetic field is very
strong. As stated in the introduction, in
this case it is sufficient to truncate the fermion propagator ( in the
absence of the $U_S(1)$ interactions) to the lowest Landau level 
(\ref{1.2}), so that we get:
\be
\tilde{S}_F^{LLL}(k) = i e^{-\frac{{\bbox {k}}^2}{eB}} \frac{1}{m + \g^0 k^0}
\left(1-i \g^1 \g^2 \right).
\label{a.1}
\ee
As we will be dealing with the lowest Landau levels only, it is expedient 
to choose the ansatz for the ``exact" propagator for the lowest 
Landau level fermions to be of the form:
\be
S_F^{LLL}(k) = i e^{-\frac{{\bbox {k}}^2}{eB}} 
\frac{1}{\Sigma(k) + A(k) \g^0 k^0}
\left(1-i \g^1 \g^2 \right).
\label{aaa.1}
\ee
Hence, following \cite{shpagin}, the gap equation for the lowest Landau
level fermion is given by
\be
\Sigma(p) =  g^2 \int \frac{d^3 k}{(2\pi)^3} e^{-\frac{{\bf k}^2}{eB}} \frac{ \Sigma(k)}{ A^2
k_0^2 + \Sigma^2(k) } D_{00}(p-k)
\label{a.3}
\ee
According to \cite{shpagin} the photon vacuum polarization gets
suppressed as $\frac{1}{\sqrt{eB}}$ at strong magnetic fields and the photons 
become almost free. Thus in the strong field limit the photon propagator 
is given in the Landau gauge by the expression:
\be
D_{\m \n}(p) = \frac{ \delta_{\m \n} - \frac{p_{\m} p_{\n}}{p^2}}{p^2} 
\frac{1}{p^2 (1+0.14037 \frac{g^2}{\sqrt{e B}})}
\label{a.3.5}
\ee

Accordingly, we have
\be
\Sigma(p) 
= {\tilde g}^2 \int d^2 {\bbox{k}} e^{-\frac{{\bbox{ k}}^2}{eB}} 
\frac{1}{ ( p-k)^2} \left(1-\frac{(p_0-k_0)^2}{(p-k)^2} \right) 
\int \frac{d k_0}{(2\pi)^3} \frac{ \Sigma(k)}{ A^2
k_0^2 + \Sigma^2(k) } ,
\label{a.4}
\ee
where ${\tilde g}^2 \equiv \frac{g^2}{1+0.14037 \frac{g^2}{\sqrt{e B}}}.$
Let us set $p$ to zero. Then,
\be
\Sigma(0) = {\tilde g}^2 \int e^{-\frac{{\bf k}^2}{eB}} {\bf k}^2
\frac{d^2 {\bf k}}{(k_0^2 + {\bf k}^2)^2} 
\int \frac{d k_0}{(2\pi)^3} \frac{ \Sigma(k)}{ A^2
k_0^2 + \Sigma^2(k) } 
\label{a.5}
\ee
For strong fields $eB \ra \infty$, we suppose that setting $\Sigma(k)
\approx \Sigma (0)$ and $A \approx 1$ yield a sufficiently good 
approximation~\cite{ian}. 
Setting ${\bbox{ k}}^2 \equiv x$, the gap equation becomes:

\be
1=  \frac{{\tilde g}^2}{ 8 \pi^2} \int dk_0 \int dx e^{-\frac{x}{eB} } 
\frac{x}{(k_0^2 + x)^2} \frac{1}{ k_0^2 + \Sigma^2 (0)}.
\label{a.7}
\ee

Assuming that the dynamically generated fermion mass is much smaller than
the external magnetic field, i.e.  $\Sigma (0) << \sqrt{eB}$, we cut off the
$x$ integration by $\sqrt{e B}$ and after the $k_0$ integration 
we get the transcendental equation:
\be
\Sigma (0) \approx\frac{ {\tilde g}^2}{4\pi} \int^{\sqrt{eB}}_{\Sigma(0)}  
dy e^{-\frac{y^2}{eB}} \left( \frac{2}{y}-\frac{3 \Sigma(0)}{y^2} 
+\frac{\Sigma(0)^3}{y^4} \right) 
\ee
The final result reads:
\be
\Sigma(0) \approx 2 {\tilde \alpha}  \ln \left(
\frac{\sqrt{eB}}{\Sigma (0)} \right)+ O(\frac{\Sigma(0)}{\sqrt{e B}}),
~{\rm where}~~{\tilde \alpha} \equiv 
\frac{{\tilde g}^2}{4 \pi}.
\label{a.8}
\ee
This equation can be solved numerically as in \cite{kostas}, 
leading to a logarithmic scaling of 
the induced fermion condensate 
with the magnetic field, 
$\Sigma (0) \sim {\rm ln}(\sqrt{eB}/{\tilde \alpha})$.

Note that the most important aspect of this type of behavior comes
from the presence of the exponential factor in the form of the free
propagator in (\ref{a.1}). However, when the external magnetic field
is weak, one has to include all the higher Landau levels as the levels
become closely spaced. Also, the wavefunctions for these levels grow
with momentum as they involve Laguerre Polynomials. Hence, one has to
work with a generic ansatz for the fermion propagator, such as (\ref{2.3}).   
However, in view of the breaking of translational invariance 
by the field-dependent phase factor in (\ref{confprop}),   
a straightforward  application of this ansatz is not possible. 
Nevertheless, as we shall discuss below, 
such an ansatz can still give qualitatively
correct predictions for the scaling of the induced condensate 
with the external field.

\section{The dynamical fermion mass in weak external magnetic fields
under  quenched approximation}

We are looking at the leading scaling behaviour 
with the magnetic field intensity 
of the 
dressed fermion propagator in the presence of an 
external magnetic field $eB$ for the case of 
weak fields $eB << \Sigma (0)$ where $\Sigma (0) \equiv m$ 
is a dynamically generated mass due to the statistical 
$U_S(1)$ interactions in the model. Obviously, since $m \propto g^2$,
whereas $g$ denotes the coupling of these interactions, the weak field
limit is achieved for relatively strong gauge interactions. 
However, as stated earlier, we are interested in the behavior of the system 
under weak magnetic fields as well and in this regime the 
Landau level decomposition is not particularly helpful.

This is a technically involved problem,
and we shall not attempt to solve it exactly in what follows.
Instead we shall make an attempt to present an
{\it approximate} treatment, which hopefully captures 
the important {\it qualitative} features of the phenomenon.
As we shall see later on, preliminary lattice calculations
will support the results obtained in this section.

The main complication arises from the fact, mentioned in 
the previous section, 
that the 
presence of an external field, perpendicular to the spatial 
plane, breaks, in addition to Lorentz invariance, also 
translational invariance in the spatial plane. 
This is apparent from the configuration space form (\ref{confprop}) 
of the 
fermion Green's function.  
The breaking of translational invariance is manifested
through the field-dependent phase factor in (\ref{confprop}).
A Schwinger-Dyson equation for $G(x,y)$ can then be obtained
as in \cite{shpagin}, but unfortunately, due to the 
form (\ref{confprop}), 
a passage to momentum space with an appropriate Fourier
transform based on a single momentum variable is not feasible.

Below we shall make a modest attempt to
calculate analytically the scaling behaviour of the 
chiral condensate with the magnetic field, at least 
qualitatively, and then compare with the lattice results. 
This is possible by adopting an ansatz for the 
fermion propagator in momentum space, $S_F^{-1}(p)$,
as if the translational invariance breaking 
phase factor in (\ref{confprop}) was absent.
Specifically, we assume that 
the fermion propagator in the presence of a {\it weak} 
external magnetic field 
can still have a Fourier transform, based on a single momentum 
variable,  
which takes the {\it approximate} form:
\be
S_F^{-1}(p)=[A_0 +A_1 \g_1 \g_2] \g_0 p_0 +C~ \bbox{\g \cdot p} 
+ [\Sigma_0 + \Sigma_1 \g_1 \g_2].
\ee
All the qualitative information
about the effects of the external field $B$ is encoded inside
the coefficients. As we shall see later on, 
by comparison with the preliminary lattice results
in the quenched approximation, 
the qualitative
features of the scaling of the magnetically-induced chiral
condensate with the external field seem to be correctly captured
by the above ansatz. It is understood, of course, that 
these results should be taken with caution and viewed 
only as preliminary. A complete lattice analysis involving 
dynamical fermions, which will settle these issues, 
is still pending.
Hopefully it will constitute the topic of a forthcoming work.
 
Under the above approximation, the Schwinger-Dyson equation for the fermion propagator in the massless limit 
assumes the form:
\be
S_F^{-1}(p) = \tilde{S}_F^{-1}(p)|_{m=0}-g^2 \int \frac{d^3k}{(2\pi)^3}
D_{\mu\nu}(p-k)\g ^\mu  S_F(k) \Gamma^\nu 
\label{sdf}
\ee
 
It is sufficient for our purposes to use the bare photon vertex 
$(\Gamma^\nu = \gamma^\nu),$ and set the wave-function renormalization to one. 
These will be justified later on.

Taking the trace in the above equation, we obtain:
\be
4 \Sigma (p) = -g^2 \int \frac{d^3k}{(2\pi)^3}
D_{\mu\nu} (p-k) {\rm Tr}\left(\g ^\mu {\tilde S}_F(k) \g^\nu \right)
\label{gap1}
\ee
where we denoted $\Sigma_0$ by $\Sigma(p).$ A more
important remark is that inside the integral we have approximated the (unknown)
fermion propagator $S_F(k)$ by the form ${\tilde S}_F(k)$ of equation 
(\ref{schwing}), which is the 
propagator in a homogeneous external magnetic field. 
One has:
\bea
&~&{\rm Tr}\left(\g ^\mu {\tilde S}_F(k) \g^\nu \right) = 
\int _0^\infty ds e^{-s (k_0^2 + \bbox{k}^2 \frac{{\rm tanh}z}{z} 
+ \Sigma ^2 )}{\rm Tr} [\g ^\mu (-\g \cdot k + \Sigma (k) )\g^\nu +  \nn \\ 
&~& i \g^\mu (-\Sigma(k) + \g \cdot k)\g _1 \g_2 \g^\nu {\rm tanh}z 
- i \g^\mu (\g_1 k_2 - \g _2 k_1 )
{\rm tanh}z (1 - i \g _1 \g_2 {\rm tanh}z)\g ^\nu ]
\label{trace}
\eea
Since we work in the Landau gauge, $D_{\mu \nu}(q)$ is given by  
$\frac{q^2 \delta_{\mu \nu}-q_\mu q_\nu}{q^4}.$ Using the following 
identities of the $\gamma$ matrices in Euclidean space:
\bea
&~& {\rm Tr}(\g _\mu \g^\mu)=-12 \nn \\
&~& (q \cdot \g)^2 = - q^2 \nn \\
&~& {\rm Tr}(\g ^\mu {\cal O} \g^\nu)D_{\mu\nu}(q)=-\frac{2}{q^2}{\rm Tr}{\cal O}
\label{identities2}
\eea
where ${\cal O}$ is any operator, one can write (\ref{trace}) in the form: 
\be
D_{\mu \nu}(p-k) {\rm Tr}\left(\g ^\mu {\tilde S}_F(k) \g^\nu \right) 
= -\frac{8}{(p-k)^2}
\int _0^\infty ds \Sigma (k)~e^{-s(k_0^2 + \bbox{k}^2 \frac{{\rm tanh}z}{z}
+ \Sigma ^2 (k))}.
\label{tracesimple}
\ee
Using (\ref{tracesimple}), (\ref{gap1}) becomes:
\be
\Sigma (p) = \frac{2g^2}{(2\pi)^3}\int _0^\infty ds \int \frac{d^3k}{(k-p)^2}
\Sigma (k) e^{-s(k_0^2 + \bbox{k}^2 \frac{{\rm tanh}z}{z}+\Sigma^2 (k))}
\label{gap2}
\ee
Setting $p=0$ and approximating $\Sigma (k) \simeq \Sigma (0) \equiv m$ 
in the integrand yields:
\be
1 = \frac{2 g^2}{(2\pi)^3} \int _0^\infty ds \int
d^3k \frac{1}{k_0^2 + \bbox{k}^2} e^{-s(k_0^2 +
\bbox{k}^2\frac{{\rm tanh}z}{z} + m^2)}.
\label{a2}
\ee
Now we use the parametrization: $k_0=k cos \theta,
~~\bbox{k}^2=k^2 sin^2 \theta,$ and get:
\be
1=\frac{g^2}{4 \pi^{\frac{3}{2}}} \int_0^\infty ds
\int_{-1}^{+1} dx \frac{e^{-s m^2}}{\sqrt{s(x^2+(1-x^2) 
\frac{{\rm tanh} z}{z})}},
\ee
with $x \equiv cos \theta.$ Then the $x$ integration can be performed:
\be
1 = \frac{1}{4 \pi^{\frac{3}{2}}} \frac{g^2}{\sqrt{e B}} \int_0^\infty d z
e^{-z \frac{m^2}{e B}} \frac{1}{\sqrt{z- {\rm tanh} z}} 
~log \frac{\sqrt{z} + \sqrt{z- {\rm tanh} z}}{\sqrt{z} - \sqrt{z- {\rm tanh} z}},
\ee
where we have set $z \equiv e B s.$
At this point we define two dimensionless variables: $\mu \equiv 2 \pi 
\frac{m}{g^2},~~~f \equiv 2 \pi \frac{\sqrt{e B}}{g^2},$ 
and write the last equation in terms
of $\mu$ and $f:$
\be
1 = \frac{1}{2 \sqrt{\pi}} \frac{1}{f} \int_0^\infty d z
e^{-z \frac{\mu^2}{f^2}} \frac{1}{\sqrt{z- {\rm tanh} z}}
~log \frac{\sqrt{z} + \sqrt{z- {\rm tanh} z}}{\sqrt{z} - \sqrt{z- {\rm tanh} z}}.
\ee
For each value of $f$ the above relation gives a corresponding value 
for $\mu.$ The result is a universal curve $\mu(f),$ which we depict in
the figure \ref{f3} as a dashed line.
This result can be translated very easily to the language of
dimensionful parameters, as well as to the lattice parameters. 
Notice that up to now there is no restriction to weak fields.

One may also derive an analytical approximation for the regime of weak
magnetic fields. Starting from eq.(\ref{a2}) we write the term
$e^{-s \bbox{k}^2\frac{{\rm tanh}z}{z}}$ in the form 
$e^{-s \bbox{k}^2} \cdot e^{-s \bbox{k}^2 (\frac{{\rm tanh}z}{z}-1)},$ 
expand the second exponential in a power series of 
$z=eBs$ for weak fields and retain the terms which are at most  
of sixth order  in $eB.$
Then the same parametrization as before is used and the integrations over
$k$ and $x$ are similarly carried out. Then the equation involving the 
integral over $z$ is replaced by:
\be
1 = \frac{g^2}{2\pi} \left[ \frac{1}{m} + \frac{1}{12}\frac{e^2B^2}{m^5}
- \frac{7}{48}\frac{e^4 B^4}{m^9} 
+\frac{773}{960} \frac{e^6 B^6}{m^{13}} + {\cal O}(e^8 B^8) \right]
\label{gap5}
\ee
For weak fields we may solve equation (\ref{gap5}) by substituting the
expansion $m=m_0 +m_2 (e B)^2 +m_4 (e B)^4 + \dots$ in (\ref{gap5}) and
equating the coefficients of equal powers of $e B$ to determine 
the coefficients $m_k.$ The resulting solution of (\ref{gap5}) is:
\be
m= \frac{g^2}{2 \pi} \left(1 + \frac{4 \pi^4 e^2 B^2}{3 g^8}
-\frac{400 \pi^8 e^4 B^4}{9 g^{16}} 
+ \frac{534208 \pi^{12} e^6 B^6}{135 g^{24} } + {\cal O}(e^8 B^8) \right),
\label{gapeq2.7}
\ee
which may also be written as:
\be
\mu= 1 + \frac{1}{12} f^4 -\frac{25}{144} f^8 
 + \frac{8237}{8640} f^{12} + {\cal O}(f^{16}).
\ee
The above relations show that for weak magnetic fields the dynamically 
generated mass is quadratic in B. For somewhat bigger magnetic fields, 
however, the quadratic behaviour is compensated by a negative
quartic contribution and the increase with the magnetic field resembles
very closely a linear dependence. 
Of course for even bigger magnetic fields higher order contributions
take over.

Although we have already depicted in figure \ref{f3} 
the full solution $\mu(f),$ 
we also plot the solution of equation (\ref{gap5}) (we note here that 
we did not actually use the solution (\ref{gapeq2.7}), 
but rather the one which results if (\ref{gap5}) is quenched to order 
$B^2$ before solving); this
is done to gain some feeling about the accuracy of the quadratic 
approximation, which will be the only possible approach in the case where
also dynamical fermions are taken into account. The approximate solution is 
the full line and is restricted in the region of small B, where it is
appropriate; it is quite good up to $\sqrt{\frac{e B}{g^2}} \approx 0.1.$
 
As we have just seen, in the quenched approximation 
one uses the free photon propagator 
as the fermion loops ( which modifies the fermion propagator ) are 
ignored. In the next section, we examine whether  the inclusion of the
photon polarization modifies  the scaling behaviour of 
the gap function with  $eB$ discussed above. 

Before closing this section two important 
remarks are in order. 
First, it should be noted that the 
presence of the translational invariance breaking phase factors in (\ref{confprop}), 
which have been ignored in the above treatment, will affect
the numerical coefficients
of the even powers of $B$ in (\ref{gapeq2.7}). 
This can be seen easily from the form (\ref{confprop}) 
by an expansion in powers of the weak field $B$ (and restriction to 
the real part, assuming hermiticity of the translational invariant
parts).  
The important issue is the sign of the various terms. 
As we shall see later on, comparison 
with the (quenched) lattice results confirms the scaling 
with $B$ given in (\ref{gapeq2.7}), thereby justifying 
the above approximate method of dealing with the problem,
at least for qualitative purposes.
Second, the above analytic treatment, leading to (\ref{gapeq2.7}),
was based on the 
approximation of replacing $\Sigma (k)$  
inside the 
integrals in the pertinent integral form of the gap equation
by a constant $\Sigma (0)=m$, the so-called
``constant-mass approximation''. This is sufficient 
for the qualitative purposes of this work,
where the main interest lies on the scaling of the induced condensate
with the magnetic field. 
It should be remarked though, that attempts to go beyond
the ``constant mass approximation'' have been made in the literature,
specifically in the context of three-dimensional QED
in the absence of external fields~\cite{koopmans}. The result is 
that the 
value of the induced mass $m$, obtained by keeping the momentum 
dependence of the gap function inside the pertinent integrals
in the Schwinger-Dyson equation, 
is half the value of the mass gap  
obtained under the ``constant mass approximation'', i.e. the 
zero-field limit ($B\rightarrow 0$) in (\ref{gapeq2.7}) 
should be $m =g^2/4\pi$. This should be taken into account 
in quantitative analyses of the phenomenon, and possible 
detailed applications 
in condensed-matter physics, 
which, however, go beyond the 
scope of the present work.

\section{Beyond the Quenched Approximation}

To take into account the contribution of internal fermion loops 
we begin with a study of the 
one-loop vacuum polarization graph in $QED_3$ 
in the case of even number of fermion 
flavors.
The polarization tensor in the one-loop approximation 
is given by :
\be
\Pi_{\mu \nu} ( p ) = - g^2 \int \frac{ d^3 k}{ ( 2\pi)^3} \tr \left(
\g_\m \tilde{S}(k) \g_\n \tilde{S}(k-p)\right)
\label{1.1}
\ee
where $\tilde{S}(k)$ is the fermion propagator in the presence of the 
external magnetic field (\ref{schwing}).
 
The calculation of the polarization tensor is straightforward.
Due to the fermionic loop, the effects of 
the translational invariance 
breaking phase factors in (\ref{confprop}) {\it cancel}, and 
one can go directly to momentum space as in the zero external
field situation. 
For our case, the photon polarization can be obtained easily by
performing a dimensional reduction of the four-dimensional result
\cite{tsai,reuter}. We end up with 
\be
\Pi_{\m \n}(p) = ( p^2 \delta_{\m \n} - p_\m p_\n ) N_0(p) + (
p_\perp^2  \delta_{\m \n} - {p_\perp}_\m {p_\perp}_\n ) N_1 (p) \equiv
p^2 P_{\m \n}N_0 (p)  + {\bbox{p}}^2 {P_\perp}_{\m \n} N_1 (p),
\label{1.4}
\ee
where $\bbox{p}^2 \equiv p_\perp^2$ with ${p_\perp}_\m =(0, p_1,p_2)$ and 
\bea
N_0 (p) =   -\frac{g^2}{8 \pi ^{\frac{3}{2}}} 
\int^\infty_0 \frac{ds}{\sqrt{s}} \int^{+1}_{-1} dv e^{-s \phi_0}
\frac{ z }{\sinh z} \left[  \cosh zv  -  v \coth z  \sinh zv \right]  \nn \\
N_1 (p) = -\frac{g^2}{8 \pi^{\frac{3}{2}}} 
\int^\infty_0 \frac{ds}{\sqrt{s}} \int^{+1}_{-1} dv e^{-s \phi_0}
\frac{2z}{\sinh^3 z} \left[ \cosh z - \cosh zv \right] - N_0(p)
\label{1.5}
\eea
with 
\bea
\phi_0 &&= m^2 + \frac{1-v^2}{4} p_0^2 + \frac{\cosh z - \cosh zv}{2z
\sinh z} \bbox{p}^2; \nn \\
z &&\equiv eBs
\label{1.5.5}
\eea
An outline of the derivation of the above-mentioned formulae is
provided in the appendix.

For weak magnetic fields, we will have $\sqrt{eB} << \Sigma (0),$  
where $\Sigma (0)$ is the dynamically 
generated fermion mass. Note that it is the opposite to the limit 
encountered in the case for the strong magnetic field \cite{shpagin,farak}. 
In the weak-field  limit, we can expand the above
functions in a power series of $z = s \Sigma^2 (0)\left( 
\frac{eB}{\Sigma ^2 (0)}\right)$ 
and take the leading and next to
leading order behavior as $z\ra 0$. 

We have, as $z \ra 0,$ the following expansions to order $e^2 B^2:$
\bea
\phi_0 = \Sigma^2 + \frac{1-v^2}{4} p^2 - \frac{z^2}{48}( 1-
v^2)^2 \bbox{p}^2 + O(z^4), \label{phi0} \\
N_0(p) =  -\frac{g^2}{8 \pi ^{\frac{3}{2}}}
\int^\infty_0 \frac{ds}{\sqrt{s}} \int^{+1}_{-1} dv e^{-s \phi_0} 
(1-v^2) \left[ 1-\frac{z^2}{6} (1-v^2) \right] + O(z^4), \\
N_1(p) =  -\frac{g^2}{8 \pi ^{\frac{3}{2}}}
\int^\infty_0 \frac{ds}{\sqrt{s}} \int^{+1}_{-1} dv e^{-s \phi_0}
\frac{z^2}{12}(-3 + 2v^2 + v^4 ) + O(z^4).
\label{1.5.55}
\eea
Note that when $z \ra 0$ the term $N_1(p)$ vanishes and we recover the
usual form for the polarization tensor. 

To simplify the integrals, we also expand the exponential
$e^{-s\phi_0}$ in a power series in $z$ so that,
\be
e^{-s\phi_0}  \sim e^{-s ( \Sigma^2 +  \frac{1-v^2}{4} p^2)} \left( 1
+ \frac{e^2 B^2 s^3}{48}(1-v^2)^2 \bbox{p}^2 \right)
\label{exponent}
\ee
This simplifies  the $s$-integrals. We then end up with:
\bea
N_0(p) = -&&\frac{g^2}{ 2\pi p}
[ (\frac{1}{2}-\frac{2 \Sigma^2}{p^2}) \sin^{-1} \kappa +
\frac{\Sigma}{p}  \nn \\
&& \qquad - \frac{2 e^2 B^2}{p^4}
\{
\sin^{-1} \kappa - \frac{2 \Sigma \kappa^2 (3+2 \kappa^2)}{3 p}
\} \nn \\
&&  \qquad + \frac{5 e^2 B^2 \bbox{p}^2}{2 p^6}
\{ sin^{-1} \kappa -
\frac{2 \Sigma \kappa^2 (15+10 \kappa^2 +8 \kappa^4)}{15 p} \} ]
\label{1.5.56}
\eea
\bea
N_1(p)=-\frac{g^2 e^2 B^2}{2 \pi p^5}
[sin^{-1} \kappa - p \frac{3 -\kappa^2+6 \kappa^4}{6 \Sigma} ]
\label{1.5.57}
\eea
where $\kappa^2 \equiv \frac{ p^2 }{4\Sigma^2 + p^2}$. Note that for physical 
processes $ 0\leq \kappa \leq 1$.  When, $eB=0$ and $\Sigma =0$ (i.e. $\kappa
=1$ )we recover the known results 
(equation \ref{2.5.5}), namely $N_0(p) = - \frac{g^2}{8 p},\, N_1(p)= 0$. 

On the other hand, in the presence of the magnetic field one can readily see
that in the limit $\Sigma \ra 0$ the function $N_1(p)$
blows up when $ \kappa \ra 1$ (i.e. the massless case ) due to the
presence of the factor $\Sigma$ in the denominator.
However, for a super renormalizable theory this seems unphysical.
A resolution to this puzzle can be provided by the generation of a
dynamical fermion mass $\Sigma$ ( however small ) in the presence of
the magnetic field. This observation points to the magnetic catalysis
even in the case for the weak fields in three-dimensional QED. 
Generation of such a mass would prevent the appearance of the
divergences.

For $\Sigma \neq 0$, $N_0(p)$ and $N_1(p)$ behaves, when 
$\Sigma << p$ and $p \ra \infty,$ as:
\bea
N_0(p) \approx -\frac{g^2}{8p}\left( 1- \frac{4 e^2 B^2}{p^4} + \frac{5
e^2 B^2 \sin^2 \theta}{p^4}\right), 
\label{1.5.575} \\ 
N_1(p) \approx -\frac{ g^2 e^2 B^2 }{2 \pi p^5} \left( \frac{ \pi}{2}
 - \frac{ 4 p}{3 \Sigma}\right)
\label{1.5.585}
\eea
where we have used the parametization $ \bbox{k}^2 = k^2 \sin^2
\theta$, as before. Hereafter we will be using the approximations 
(\ref{1.5.575}) and (\ref{1.5.585}) which are
 justified in the weak-field case.

As in the case with the quenched treatment let us proceed to get the
gap equation. Again we shall adopt the approximate 
qualitative approach 
of the previous section as regards the effects 
of the translational invariance breaking phase 
factors in (\ref{confprop}) due to the 
presence of the external field.
To take account of the photon propagator, 
we can invoke the large-N argument 
to sum up the photon propagator in
the ladder approximation :
\be
D_{\m \n} (p) = D^0_{\m \n }(p) + D^0_{\m \kappa}(p) \Pi_{\kappa \rho}(p)
D^0_{\rho \n}(p)  + \cdots
\label{1.6}
\ee
where $D^0_{\m \n}$ is the free  photon propagator. 
To facilitate our calculations let us use the Landau gauge for the 
zeroth order propagator, so that 
\be
D^0_{\m \n}(p) = \frac{ \delta_{\m \n}- \frac{ p_\m p_\n }{p^2}}{
p^2}
\label{1.7}
\ee
Using the algebraic properties of the projectors
\bea
P_{\m \n} (p) P_{\n \rho} (p) = P_{\m \rho}(p) \nn \\
P_{\m \n}(p) {P_{\perp}}_{\n \rho}(p_\perp) =  {P_{\perp}}_{\n
\rho}(p_\perp) \nn \\
 {P_{\perp}}_{\m \n}(p_\perp)  
{P_{\perp}}_{\n \rho}(p_\perp) =  {P_{\perp}}_{\n \rho}(p_\perp)
\label{1.8}
\eea
we can sum the series in (\ref{1.6}) to get
\be
D_{\m \n} (p) = \frac{1}{ p^2(1 - N_0(p)) } \left[ P_{\m \n} + 
{P_\perp}_{\m \n }\frac{ N_1(p)\frac{\bbox{ p}^2}{p^2}}{(1- (N_0(p) + 
\frac{\bbox{p}^2}{p^2} N_1(p))} \right]
\label{1.9}
\ee
To go beyond the case of quenched approximation which we discussed 
in the previous section, we need to include the polarization effects 
in our analysis treatment.  To perform this improvement we replace the
photon propagator in (\ref{gap1}) by the ladder resummed one, given
by (\ref{1.9}). For the fermion propagator we proceed as we did previously 
for the quenched case: 
starting from equation (\ref{trace}) we expand 
the term $e^{-s \bbox{k}^2\frac{{\rm tanh}z}{z}}$ and get the expression 
$e^{-s \bbox{k}^2} \cdot e^{-s \bbox{k}^2 (\frac{{\rm tanh}z}{z}-1)}.$
Then we expand the second exponential in powers of
$z=eBs,$ retain the terms which are at most
of second order in $eB$ and integrate over s.
Taking traces over the Dirac $\g$ matrices we finally get:
\bea
\Sigma(p) =&& g^2 \int \frac{d^3 k}{(2\pi)^3} \frac{ \Sigma(k)}{
(k_0^2 +  \bbox{ k^2} + \Sigma^2(k) )} \left(1 + 
\frac{2 e^2 B^2 \bbox{k}^2}{(\Sigma^2 + k^2)^3}\right)
\frac{1}{ (p-k)^2(1 - N_0(p-k)) } \times \nn \\
&& \qquad  \left[ 2 + \frac{ N_1(p-k)\frac{
(\bbox{p-k})^2}{(p-k)^2}}
{(1- (N_0(p-k) + 
\frac{(\bbox{p-k})^2}{(p-k)^2} N_1(p-k))} \right] 
\label{1.10}
\eea
Let us now set $p =0$ and as in the case of the quenched aprroximation
let us make the substitution $\Sigma(k) \simeq \Sigma(0) \equiv m $ to get
\bea
1 = && g^2 \int \frac{d^3k}{(2\pi)^3} 
\frac{ \left(1 + 
\frac{2 e^2 B^2 \bbox{k}^2}{(m^2 + k^2)^3}\right)}{
(k_0^2 +  \bbox{ k^2} + m^2 )}
\frac{1}{ k^2(1 - N_0(k)) } \times \nn \\
&& \qquad  \left[ 2 + \frac{ N_1(k)\frac{
\bbox{k}^2}{k^2}}
{(1- (N_0(k) + 
\frac{\bbox{k}^2}{k^2} N_1(k))} \right]
\label{1.11}
\eea
Now as  $k^2 = k_0^2 + \bbox{k}^2$ and we can use the parametrization
$ k_0 = k \cos \theta $ and $ \left|\bbox{k}\right| 
= k \sin \theta $ .  Let us write, using (\ref{1.5.575}), 
\be
N_0(p) = E(p) + F(p) \sin^2 \theta
\label{1.11.5}
\ee
where $E(p) = - \frac{g^2}{8 p}( 1- \frac{4 e^2 B^2}{p^4})$ and $F(p)=
-\frac{5 g^2 e^2 B^2}{8 p^5}$. Accordingly,
we can rewrite (\ref{1.11}) as 
\bea
1 = &&g^2 \int \frac{ dk}{(2\pi)^2} 
\frac{1}{ (k^2 + m^2)}\times \int^\pi_0 d \theta \frac{\sin \theta}
{(1 - E(k)- F(k)
\sin^2 \theta ) } \nn \\    
&& \times \left[ 2 + \frac{ N_1(k) \sin^2 \theta}
{(1- E_0(k) -(F(k)+ N_1(k) )\sin^2 \theta) } \right]
\left(1 + \frac{2 e^2 B^2 k^2 \sin^2 \theta}{(m^2 + k^2)^3 }\right)
\label{1.12}
\eea
The angular integral can be performed by making a change of
variables $y=\cos \theta,$ so that we end up with: 
\bea
1 =  && \frac{g^2}{\pi^2} \int  dk 
\frac{1}{(k^2 + m^2)} \left[ \frac{1}{F(k)} \left( \left\{ 1+ \frac{2e^2 B^2
k^2 (b^2 +1)}{(k^2 + m^2)^3 } \right\} \frac{1}{b}
\tan^{-1}(\frac{1}{b}) - 2\frac{e^2 B^2 k^2}{(k^2 + m^2)^3} \right)  \right.
\nn \\
&& + 
\frac{ e^2 B^2 k^2 N_1(k)}{ (k^2+ m^2 )^3 F(k) ( F(k) + N_1(k))} -
\frac{1}{2(1-E(k))} \left( \left\{ 1+ \frac{2(b^2+1)e^2 B^2 k^2}{(m^2 +
k^2)^3} \right\} \frac{b^2 +1}{b} \tan^{-1} (\frac{1}{b}) \right. \nn \\
&&\qquad -  \left. \left. \left\{ 1+ \frac{2(a^2+1)e^2 B^2 k^2}{(m^2 +
k^2)^3} \right\} \frac{a^2 +1}{a} \tan^{-1} ( \frac{1}{a}) \right) \right],
\label{1.13}
\eea
where $b^2 \equiv \frac{1-E - F}{F}$ and $a^2 \equiv
\frac{1-E-F-N_1}{N_1 + F}$.

However, this equation is difficult to handle, so we prefer instead 
to expand equation (\ref{1.12}) 
in powers of $e^2 B^2$ {\em before} doing the angular integration. 
After some rather tedious algebraic manipulations we end up with:
\be
\frac{e B}{g^4} \equiv \frac{f^2}{4 \pi^2} = \sqrt{\frac{1-A_0}{T_1+T_2}}.
\ee
The quantities appearing in the right hand side are given by the following
expressions:
\be
A_0 = \frac{2}{\pi \mu} \int_0^\infty dx \frac{x^2}{(x^2+1) [x^2+h(\mu,x)]},
\ee
\be
T_1 = \frac{128 \pi^3}{3 \mu^5}
\int_0^\infty dx \frac{x^4}{(x^2+1)^4 h(\mu,x)},
\ee
\be
T_2 = \frac{128 \pi^3}{9 \mu^6} \int_0^\infty dx \frac{12 x^4 + x^6}{
(4+x^2)^3 (x^2+1) [h(\mu,x)]^2}.
\ee
In the above expressions we have used the notations $f \equiv 2 \pi 
\frac{\sqrt{e B}}{g^2}$ and $\mu \equiv 2 \pi \frac{m}{g^2},$ 
already used in the previous section; moreover, we have employed 
the expression:
$$h(\mu,x) \equiv \frac{1}{\mu} 
\left[ 1+\frac{x^2-4}{2 x} sin^{-1} 
(\frac{x}{\sqrt{x^2+4}}) \right].$$

Numerical computation of the integrals yields $\frac{m}{g^2}$ as a function
of $\frac{\sqrt{e B}}{g^2},$ which can be used to produce the lower 
branch in figure \ref{f3}. This represents the solution of the SD equations 
in the region of small magnetic fields. (Note that only the small-$B$ part
of the curve is depicted). We see that the dynamically 
generated mass in this case is substantially smaller than in the previous 
section. Presumably this reflects the fact that, due to the Pauli principle, 
the condensate tends to decrease. We note that in the quenched case the 
back reaction of the fermions is not really felt, so this fact has no 
consequences in that case. 

\section{Lattice formulation}

We now proceed with a description of the lattice formulation of the problem. The
lattice action is given by the formulae given below.

\bea
&~&S =\frac{\beta_G}{2} \sum_{x,\mu, \nu} F_{\mu \nu}(x) F^{\mu \nu}(x) 
+ \sum_{n,n^\prime} {\overline \Psi}_n Q_{n,n^\prime} \Psi_{n^\prime}
\eea
$$
F_{\mu \nu}(x) \equiv a^S_\mu(x)+a^S_\nu(x+\mu)-a^S_\mu(x+\nu)-a^S_\nu(x)
$$
$$
Q_{n,n^\prime} = \delta_{n,n^\prime}-K \sum_{\hat \mu} 
[\delta_{n^\prime,n+\hat \mu} (r+\gamma_{\hat \mu}) U_{n {\hat \mu}} 
V_{n {\hat \mu}} + \delta_{n^\prime,n-\hat
\mu} (r-\gamma_{\hat \mu}) U_{n-\hat \mu, \hat \mu}^\dagger 
V_{n-\hat \mu, \hat \mu}^\dagger].
$$
The indices $n,~n^\prime$ consist actually of three integers each, 
$(n_1,~n_2,~n_3),$
labeling the lattice sites, while $\mu$ denotes directions.
$r$ is the Wilson parameter, $K$ the hopping parameter, 
$U_{n {\hat \mu}} \equiv
e^{i g a \alpha^S_{n {\hat \mu}}},~~V_{n {\hat \mu}} 
\equiv e^{i e a A_{n {\hat \mu}}}.$ 
$\alpha^S_{n {\hat \mu}}$ represents the statistical gauge
potential and $A_{n {\hat \mu}}$ the
external electromagnetic potential. $\beta_G \equiv \frac{1}{g^2 a}$
is related to the statistical gauge coupling constant
in the usual way. On the other hand, we denote by $e$ the 
dimensionless electromagnetic
coupling constant of the external electromagnetic field. 
In our treatment we will use na\"ive fermions, so we set $r=0.$
Initially we will consider a 
homogeneous magnetic field; thus one should construct a lattice
version of the homogeneous magnetic field. This has already been 
done before in \cite{damhel} in connection with the abelian Higgs model.
We more or less follow them, but follow a slightly different prescription,
which we describe below \cite{lattice}.

Since we would like to impose an external homogeneous
magnetic field in the (missing) $x_3$ direction, we
choose the external gauge potential in such a way that the plaquettes in
the $x_1 x_2$ plane equal B, while all other plaquettes equal zero.
One way in which this can be achieved is through the choice:
$A_3(n_1, n_2, n_3)=0, {\rm for~all }~n_1,~n_2,~n_3,$ and
\bea
A_1(n_1, n_2, n_3) =-\frac{B}{2} (n_2-1), n_1 \ne N, & 
A_1(N, n_2, n_3)  =-\frac{B}{2} (N+1) (n_2-1),\\
A_2(n_1, n_2, n_3) =+\frac{B}{2} (n_1-1), n_2 \ne N,&
A_2(n_1, N, n_3) =+\frac{B}{2} (N+1) (n_1-1).
\eea
where $N^3$ is the number of points on the (cubic) lattice.  
It is trivial to check out that all plaquettes starting at $(n_1,n_2,n_3),$
with the exception of the one starting at $(N,N,n_3),$ equal $B.$ The latter 
plaquette equals $(1-N^2) B = B -(N^2 B).$ One may say that the flux
is homogeneous over the entire $x_1x_2$ cross section of the lattice and
equals $B.$ The additional flux of $-(N^2 B)$ can be understood by the
fact that the lattice is a torus, that is a closed surface, and the 
Maxwell equation
$\bbox{\nabla \cdot B} =0$ implies that the magnetic flux through the lattice
should vanish. This means that, if periodic boundary conditions are
used for the gauge field, the total flux of any configuration should be 
zero, so the (positive, say) flux $B,$ penetrating
the majority of the plaquettes, will be accompanied by a compensating
negative flux $-(N^2 B)$ in a single plaquette.
This compensating flux should be ``invisible", that is it should have no
observable physical effects. This is the case if the flux is an integer
multiple of $2 \pi: N^2 B =m 2 \pi \to B=m \frac{2 \pi}{N^2},$ 
where $m$ is an integer. Thus we may say (disregarding the ``invisible" 
flux) that the magnetic field is homogeneous over the entire cross section
of the lattice.\footnote{To check this translational invariance we 
measured the fermion condensate at every point in the $x_1 x_2$ plane. 
The results were the same at all points within the error bars, 
confirming homogeneity.} The integer $m$ may be chosen to lie in the 
interval $[0, \frac{N^2}{2}],$ with the
understanding that the model with integers $m$ between $\frac{N^2}{2}$ 
and $N^2$ is equivalent to the model with integers 
taking on the values $N^2-m,$ which are among the
ones that have already been considered. It follows that the magnetic field 
strength B in lattice units lies between 0 and $\pi.$ The physical 
magnetic field $B_{phys}$ is
related to $B$ through $B=e \alpha^2 B_{phys},$ and the physical field may
go to infinity letting the lattice spacing $\alpha$ go to zero,
while $B$ is kept constant.

An important remark is that the magnetic field is not allowed to be too
big in lattice units, since then the perturbative expansion of the expressions 
$e^{i e a A_{n \mu}}$ would yield significant $B^2,~B^3,\dots$ 
contributions with the accompanying vertices, in addition to the
desirable terms which are linear in $B.$ A trivial estimate of the critical 
field strength is obtained from the demand that the cyclotron radius 
corresponding to a given magnetic field should not be less than (say) two
lattice spacings. This trivial calculation yields $B < \frac{\pi}{8}.$
Of course the above limitations apply strictly only to the case where the
statistical gauge field has been turned off; in the ``interacting" case, 
one does not really know whether there exists a critical magnetic field, 
after which discretization effects are important.
With this remark in mind, we depict in the figures of the following sections 
the results for the whole range of the magnetic field, from 0 to $\pi.$

For the fermion fields we used antiperiodic boundary conditions in 
the time direction and ``fixed"
boundary conditions in the spatial directions; the latter boundary 
conditions mean that we consider
fermion fields vanishing on the boundaries.

\section{Lattice Results}
\subsection{Zero Temperature Results}
We are now going to present the results pertaining to the $T=0$ case.
The first set consists of measurements of the fermion condensate 
versus the magnetic field for a $16^3$ lattice in the strong coupling 
regime for the statistical gauge field $(\beta_G = 0.10)$ for three 
values of the bare mass (figure \ref{f4}). Before going on with the specific 
features of this results, let us remark that to facilitate comparison 
with the analytic results we measured the magnetic field in 
units of its maximal value: thus we used the parameter $b,$ defined 
by: $b \equiv \frac{B}{B_{max}} = 
\frac{ e B_{phys} a^2}{e B_{phys} a^2|_{max}}.$
Since $B_{max} = \pi,$ as explained previously, we get: 
$b = \frac{B}{\pi}$ and $b$ runs from 0 to 1.
We see in figure \ref{f4} that for all three masses the plot consists of
two parts with qualitatively different behaviour. For $b$ smaller than
about 0.3 we find a dependence of the condensate on the external
magnetic field, which is nearly linear, however, in view of the 
analytical results obtained in section 5 about the quenched case, we 
understand that we see the quadratic behaviour found there; however, there 
is a negative quartic term coming into the game, as we also saw, and this 
``straightens out" the quadratic curve and makes it almost linear.  
For big magnetic fields we find points that
could possibly be fitted to a logarithmic type of curve. The 
logarithmic dependence 
$$
\frac{\Sigma(0)}{\alpha} \simeq ln \left [\frac{\sqrt{e B}}{\alpha}\right ],
~~\alpha \equiv \frac{g^2}{4 \pi},
$$
has been found \cite{fm} by an approximate solution of the 
Schwinger-Dyson equations in the regime of strong magnetic fields. 
We have included such a logarithmic fit for $m=0.050$
in figure \ref{f4}. In addition, for this mass some points in the 
intermediate region are included. They show a smooth interpolation 
between the two regions. 
Thus in both the strong and weak magnetic field regimes we find a 
nice qualitative agreement of the analytical solutions 
with the Monte Carlo results. In the figure we have also included the
extrapolation to the zero mass limit.

The magnetic field has been characterized as ``strong" or ``weak" through
its comparison with the fine structure constant 
$\alpha \equiv \frac{g^2}{4 \pi}$ and the dynamically generated mass
$m \simeq \Sigma(0).$ Since figure \ref{f4} contains the strong coupling data, 
it would be interesting to explore the
fate of the whole picture shown in figure \ref{f4} as the gauge 
coupling $g$ moves
away from the strong coupling regime. One would na\"ively expect that 
the magnetic fields will be more easily characterized as ``strong", as
compared to the smaller and smaller scale set by the gradually weaker
coupling constant. Thus, the almost linear part should be restricted to
the very small magnetic fields and eventually disappear. This is what 
one may see in figure \ref{f5}, which is similar to figure 
\ref{f4}, the only difference 
being that the gauge coupling constant is now in the intermediate
coupling regime, rather than the strong coupling of figure \ref{f4}. We
see that the almost linear part is now restricted in the region between
$b=0$ and $b \approx 0.12-0.15.$ We may also have a semi-quantitative 
estimate of the new ``critical" magnetic field $b_c,$ defined as the
maximum $b$ which fits into the almost linear behaviour. Inspired by
the inequality $e B << (\frac{g^2}{4 \pi})^2,$ let us suppose
that $e B_c = x (\frac{g^2}{4 \pi})^2,$ with $x$ a very small number;
we also make the further assumption that $x$ does not depend on $g.$ 
We will check crudely whether this assumption is  reasonable given our 
results. Converting everything to lattice units,
we find that $b_c = \frac{x}{16 \pi^3} \frac{1}{\beta_G^2}.$  
From this we infer $b_{c_2}=b_{c_1} (\frac{\beta_{G_1}}{\beta_{G_2}})^2.$
Using $\beta_{G_1}=0.10,~\beta_{G_2}=0.15$ and $b_{c_1}=0.3,$ we find
for $b_{c_2},$ the value 0.13, which is surprisingly close to the
value given by the data of figure \ref{f5}. 
Of course, the $\frac{1}{\beta_G^2}$
dependence of $b_c$ tells us that the weak field behaviour will be 
even more suppressed as we move towards the weak gauge coupling; this is what
we have seen in direct simulations in this regime. Thus, it is plausible
that the dependence of the ``critical" magnetic field has a 
$\frac{1}{\beta_G^2}$ dependence.

We now make contact with the results of \cite{lattice}, where we 
studied the model 
with the statistical gauge field turned off. We had found there that 
for big enough $b$ the condensate stopped showing a monotonous
increase with $b,$ at $b=0.5$ it had a local minimum and then had a
succession of maxima and minima, up to $b=1.$ Moreover, there was 
a spectacular volume dependence. One expects, of course that this 
``free" case 
will be reached for big enough $\beta_G.$ In figure \ref{f6} 
we show the results 
for $\beta_G=0.5$ and $\beta_G=1.0$ for various volumes. For 
$\beta_G=0.5$ the curve shows the first sign of 
non monotonous behaviour 
at $b=0.5,$ while at $\beta_G=1.0$ the succession of maxima and minima is
clear. However, there is no detectable volume dependence, so we can 
be sure that, even at this large $\beta_G,$ the limit of switching the
gauge field off has not yet been reached; it will presumably be reached
for even bigger values of $\beta_G.$ One should add that in the ``free" 
case the role of the bare mass is very important, since it is eventually
the only source of mass generation. This is at the root of the large
volume dependence showing up in the ``free" case: at fixed volume  
the condensate goes over to zero for vanishing bare mass. In the 
full model, though, the interaction with the gauge field generates a 
dynamical mass, independently from the value of the bare mass. This is why
in the ``interacting" case the volume dependence is small, permitting a 
smooth transition to the thermodynamic, as well as to the massless, limit.

The simulations are done at finite values for the (bare) mass; the 
massless limit is taken by extrapolating the results for several 
bare masses to the limit $m \to 0.$ Figure \ref{f7} shows the process of 
this extrapolation for three values of the gauge coupling constant. 
The external magnetic field has been set equal to a typical value 
$(b=0.188);$
the picture is similar for all values of the magnetic field strength.
For $\beta_G=0.10,$ which lies in the strong coupling region, the 
extrapolation is linear with negative slope. This line is pointing 
to a relatively big value for the condensate in the chiral limit. 
For somewhat weaker coupling $(\beta_G=0.20),$ the curve is still 
a straight line, but the slope is positive and it points to a smaller
value at $m \to 0.$ In both of these cases the mass dependence is not
very pronounced, because it is the strong gauge coupling which dominates
in the formation of the condensate. In the third case $(\beta_G=0.35),$ 
which lies in the weak coupling, one can no longer fit to a straight line;
a quadratic fit has proved necessary for all $\beta_G$ values smaller 
than $0.2.$

Figure \ref{f8} contains the zero mass limit of the condensate (obtained
through the procedure illustrated in figure \ref{f7}) versus $\beta_G,$ 
for four values of the external field. We observe that in the 
strong coupling region the b-dependence of the condensate is rather 
weak; on the 
contrary, at weak coupling, the external field is the main 
generator of the condensate, and we find an increasingly big 
b-dependence, as we move to large $\beta_G.$ 
Note that the biggest value of $b$ we have used in systematic 
measurements, such as the ones in figure \ref{f8}, is 0.3. This takes 
into account that for larger values of $b$ the function of the
condensate stops being monotonous for large $\beta_G,$ as may be 
seen on figure \ref{f6}. Thus we have restricted our study to a $b$ region
which is safe for all values of the coupling. From this preliminary 
quenched study we conclude that a non-vanishing value for 
$<{\overline \Psi} \Psi>$ develops for this small volume even at weak 
coupling in the presence of an external magnetic field. We have not 
tried to study systematically the approach to the continuum limit.

\subsection{Finite Temperature Results}

We expect that the fermion condensate, generated by any mechanism 
(explicit mass term, gauge interactions, external fields) should
vanish at high temperatures. This is the study we now turn to: we 
employ asymmetric lattices and consider the behaviour of the 
condensate versus $\beta_G.$ Before describing the behaviour of
the condensate, let us first see the $\beta_G$ dependence of the 
Wilson line.

Figure \ref{f9} depicts the Wilson line versus $\beta_G$ for lattices 
of temporal size $N_\tau = 4$ and various spatial volumes. We 
observe in the figure not only the decrease of the Wilson line 
with increasing spatial volume, but also the fact that initially this 
quantity is almost independent of $\beta_G,$ but at $\beta_G 
\simeq 0.25$ its dependence on $\beta_G$ starts showing up.
It is important that this value of $\beta_G$ is independent from 
the spatial volume.

In figure \ref{f10} we show the relationship between Wilson lines on lattices 
with $N_\tau = 2$ versus $N_\tau = 4$ and $N_\tau = 6.$ 
The result for the $16^2 * 6$
lattice lies ``below" the result for the $16^2 * 4.$ This is due to 
the fact that the former lattice is closer to the zero temperature
(symmetric) lattices, as compared to the latter. The value of 
$\beta_G,$ above which there is $\beta_G$ dependence is
substantially bigger for $N_\tau = 6$ than for $N_\tau = 4$ 
or $N_\tau = 2.$ The Wilson line for the $N_\tau = 2$ case 
approaches an asymptotic value for large $\beta_G.$ This is 
not very obvious in the other two cases, because they lie farther 
from the infinite temperature limit. Also in this case the 
statistical fluctuations are very large, resulting in big errors.
With this in mind we have put in the figure only the errors for 
the case $N_\tau = 2.$.

Figure \ref{f11} contains the zero (bare) mass extrapolations for the 
condensate as a function of $\beta_G.$ The external magnetic
field is set to $b=0.1.$ The uppermost curve
contains the results for a symmetric lattice $(16^3.)$ It is 
easily seen that it is a quite smooth curve and presents no apparent 
discontinuities of any sort. The data for the asymmetric lattice
$16^2 * 4$ follow the ones of the symmetric lattice at strong coupling;
in the weak coupling region the condensate for the asymmetric lattice 
appears substantially smaller than its $16^3$ counterpart. This
is what one should explain on account of the symmetry restoration 
scenaria at finite temperatures. The $16^2 * 4$ data 
can be described by two branches, one containing the strong and the 
other the weak coupling results; the two branches join at about 
$\beta_G = 0.4,$ but their slopes do not coincide. There is a
discontinuity at this value of $\beta_G,$ which we interpret
as the symmetry restoring transition at finite temperature. 
On the same figure we have put the results for a lattice of 
bigger spatial volume $(24^2 * 4),$ away from the ``critical" 
$\beta_G = 0.4$ value. These data do not differ 
substantially from the ones for $16^2 * 4.$ 

Figure \ref{f12} contains results similar to the ones of figure \ref{f11}, but the
value of the external magnetic field differs: $b=0.305.$
The same basic picture appears here, as well: we may again spot the
discontinuity at $\beta_G =0.4$ for $N_t = 4.$ 
In addition to the data of figure \ref{f11} 
we have put the data for a $24^2 * 6$ lattice, which is expected to 
lie closer to zero temperature. The data are smoother than the ones
for $16^2 * 4$ and they lie much closer to the $16^3$ results; this 
makes it more difficult than before to spot a sudden change in slope; 
however, this change is present even in this case. The new element 
here is the data for $N_t=2.$ The change in slope here is very
pronounced and substantiates our claim that we have a phase 
transition around $\beta_G =0.4.$

Since we now have data for several asymmetric lattices, we are 
in a position to show the temperature dependence of the condensate.
This is done in figure \ref{f13}, in the weak coupling regime, for two 
values of the magnetic field. The zero mass extrapolation of the
results has been used and the temperature in lattice units is 
$\frac{1}{N_t},$ as usual. We observe the fall of the condensate at
high temperatures, which is more dramatic for the smallest value of 
$b.$ This figure is of the same qualitative form with figure 2 of 
\cite{lattice}, which was derived analytically for the case where we had
no statistical gauge field at all.

In figure \ref{f14} we show the time evolution of the condensate for two
values of the magnetic field at weak coupling for a symmetric lattice.
The important feature here is the very small magnitude of the statistical 
fluctuations, resulting in relatively small errors.

The situation in figure \ref{f14} changes in the asymmetric lattices and the
results of figure \ref{f14} should be contrasted against the ones in figures 
\ref{f15} and \ref{f16}. In figure \ref{f15} we have exactly the same 
parameters as in figure \ref{f14},
however now we have a $16^2*4$ lattice. It is evident that the
fluctuations have grown about one order of magnitude larger.

The statistical fluctuations grow even larger for the $16^2*2$ lattice, 
whose results are shown in figure \ref{f16}. Moreover, this figure gives a 
feeling of the way the average of the condensate is approaching zero
at high enough temperatures. The outcome spends most of its time at 
small values and has some exceptional big spikes from time to time;
these latter become more and more rare as the spatial volume increases.

\subsection{Non-uniform Magnetic Field}

In the previous sections we have 
considered the case of a uniform external magnetic 
field. There is however potential physical 
interest in the effects of non-uniform fields, which 
become important
in case the above model has relevance to the physics of  
high-temperature superconductors. Indeed, it will be of interest
to examine the effect of electromagnetic vortex lines 
on the induced fermion (holon) condensate 
at the nodes of $d$-wave superconducting 
gaps~\cite{kostas}. A simple-minded model for such vortex lines could be 
that of flux tubes of magnetic field. 
The full problem would be to take into account interactions 
among the flux tubes, which could shed light 
also in the confining aspects of the gauge groups in three dimensions.
It is only by lattice methods that one may treat the problem, in
view of the very big computational difficulties in the analytical approach. 
In this first treatment of the problem we switch off the fluctuating 
statistical gauge field and consider the response of the fermions to 
the background field only. 
A full treatment of the problem, including the 
statistical gauge interactions is left for the future.

Let us describe the technical procedure to construct a non-uniform
magnetic field on the lattice. We will consider 
$M \times M$ plaquettes parallel to the $x_1 x_2$ plane, 
around the center of the lattice, which 
will be penetrated by magnetic flux equal to $B$ each. The remaining
plaquettes will not carry any flux. Then we are going to measure the
condensate at the center $(\frac{N}{2},\frac{N}{2},\frac{N}{2})$ 
and along a line passing through it and consisting of the sites 
$(\frac{N}{2},\frac{N}{2}+1,\frac{N}{2}),$ 
$(\frac{N}{2},\frac{N}{2}+2,\frac{N}{2}), 
\dots,$ $(\frac{N}{2},N,\frac{N}{2}).$

The fact remains that the total flux through the lattice should be
zero, because of $div {\bf B} =0.$ Thus, for each flux B penetrating
a given plaquette, there should be an opposite flux somewhere else 
in the lattice. To construct the magnetic field that we mentioned 
above, we followed the strategy to build it up plaquette by plaquette
taking care that we put the compensating opposite flux through the
plaquette starting at the point $(N,N,n_3).$ 
If we already have a configuration of gauge potentials on a lattice, 
the procedure to add a plaquette of flux $B$ at the plaquette at 
$(N_1,N_2,n_3)$ (with the corresponding compensating flux $-B$ at 
$(N,N,n_3)~),$ consists of adding to the preexisting links the 
quantities denoted by $\Delta A_k$ below.
$\Delta A_3(n_1,n_2,n_3)$ is set to zero for all values of the integers
$n_1,n_2,n_3.$ $\Delta A_1$ and $\Delta A_2$ are also set to zero, 
except for the links where an explicit different statement is made.
For the plaquette starting at the site $(N_1,N_2,n_3), N_1 \ne N, N_2 \ne N,$
we choose:
$$
\Delta A_1(N_1,n_2,n_3)= -B,~~ n_2=N_2+1,N_2+2,\dots,N,~~n_3=1,\dots,N
$$
$$
\Delta A_2(n_1,N_2,n_3)= +B,~~ n_1=N_1+1,N_1+2,\dots,N,~~n_3=1,\dots,N
$$
For $N_1=N$ we impose:
$$
\Delta A_1(N,n_2,n_3)= -B,~~ n_2=N_2+1,N_2+2,\dots,N,~~n_3=1,\dots,N
$$
$$
\Delta A_2(n_1,n_2,n_3)= 0~~ {\rm everywhere,}
$$
while for $N_2=N:$
$$
\Delta A_1(n_1,N,n_3)= 0~~{\rm everywhere},
$$
$$
\Delta A_2(n_1,N,n_3)= B,~~ n_1=N_1+1,N_1+2,\dots,N,~~n_3=1,\dots,N
$$
In the following we consider the model with the statistical gauge
field turned off. We start with vanishing gauge potentials 
everywhere on the lattice, go through the plaquettes in the
central region and add the above $\Delta A_k$ quantities to the
corresponding links. In this way
we end up with the flux $B$ in the central plaquettes and the compensating
flux for all the plaquettes at $(N,N,n_3).$ The flux through these 
latter plaquettes should be ``invisible", as explained in section , 
so B must take the values $\frac{2 \pi}{M^2} n,~n=0,1,\dots,\frac{M^2}{2}.$

In figure \ref{f17} we show the results for a central region of non-vanishing 
flux of extent $6 \times 6.$ More specifically, for the $16^3$ 
lattice we have been using, the region with constant non-zero
flux contains the plaquettes starting at $(n_1,n_2,n_3),$ with 
$6 \le n_1 \le 11$ and $6 \le n_2 \le 11,$ while $n_3$ takes all values.
Note that nothing depends on the value of $n_3.$
The uppermost curve in the figure depicts the result for 
the condensate at the site $(9,9,9).$ The remaining
curves represent the corresponding results for the sites $(9,12,9),$
and $(9,16,9).$ The curves with $n_2=10,~n_2=11$ are quite similar to the 
$n_2=9$ curve. The first substantial change 
takes place at the site $(9,12,9),$ which lies exactly on the boundary
of the above region. The remaining curves dive together to a value 
which is accounted for by the explicit mass term and has very little to do
with the external magnetic field. Thus, we find a drop in the 
condensate value taking place exactly on the boundary of the central region. 

To visualize the fall of the condensate on the boundary, we fixed the $b$ 
parameter to 0.111 (a typical value) and plotted the value of
the condensate along a straight line passing from the center of the lattice.
We find the symmetric bell-shaped plot shown in figure \ref{f18}.

\section{Conclusions}

In this work we have studied in detail, 
by means of analytic and lattice methods,
the phenomenon of magnetic catalysis 
in even-flavour $QED_3$, namely the 
magnetic induction of a chiral-symmetry breaking 
fermion condensate as a result of the influence of an external 
magnetic field. We have shown that 
the scaling behaviour of the induced condensate 
with the external field varies according to the strength
of the latter. In the weak-field regime, there is 
a quadratic increase of the condensate with increasing 
external field, to be contrasted with the logarithmic 
scaling behaviour in the regime of strong external magnetic fields. 
However, it seems that the 
transition from weak to strong fields
is smooth, at least as far as the 
induced condensate is concerned, and we would characterise it 
as a {\it cross-over} rather than a phase transition 
at some critical value of
the external field. This constitutes a prediction of the gauge theory, 
and it may be tested in experiments of relevance to 
high-temperature superconducting materials. 
It would be interesting to repeat the (lattice) computations 
for the case of four-fermion contact interactions to check on this 
behaviour. This would differentiate between the two models as possible 
candidates for the {\it nodal} spin-charge excitations in $d$-wave 
high-temperature superconductors~\cite{kostas,ssw}.

It should be stressed that the analytic methods that 
lead to this scaling are approximate, and should be considered
only as giving a qualitative treatment of the
phenomenon. The important effect of the external field
is the breaking of 
translational invariance on the spatial plane, 
and this leads to technical complications in 
solving the pertinent Schwinger-Dyson equations in the case 
of weak magnetic fields, mixing configuration and momentum
space integrals. To bypass this problem,  
we adopted an ansatz for the fermion 
propagator in the presence of a weak 
external field, which 
although maintains formally a translational invariant look
(in the sense of its being expressed in terms of a Fourier
transform depending on a single momentum variable), however 
it incorporates the effects of the magnetic 
field in the pertinent coefficients. Comparison 
with the quenched lattice results showed that 
the predicted scaling of the induced condensate with the
magnetic field is (qualitatively) captured by this ansatz.

In addition to the uniform external field case, we have also 
presented preliminary quenched lattice results 
in the case of flux tubes of magnetic field. This situation might also 
be of relevance to realistic situations in high-temperature 
superconductors, as being related to the effects of electromagnetic 
vortex lines on the opening of a fermion 
gap at the nodes of the superconductor, within the context of the gauge theory
approach~\cite{kostas}. 
Our results in the non-uniform magnetic field case have indicated 
that the fermion chiral condensate is non zero and 
scales with the magnetic field of 
the flux tube at the core of the latter, but decays very fast outside
the tube. Our considerations did not properly take into account 
interactions among flux tubes. The latter is an important issue, 
which might also bear some relation with the issue of confinement of the 
three-dimensional theory. We expect that a proper treatment of this 
problem will become 
available only upon the use of dynamical fermions on the lattice. 

Another important issue we would like to address for future work is the 
computation of thermal conductivities in the context of the 
model of section 2, used in our simulation of the physics 
of planar high-temperature superconductors. As discussed in \cite{lattice2}
there are scaling differences of the thermal conductivity 
between the gauge ($QED_3$) and four-fermion models, which would 
be important to analyse in detail in the context discussed in this work
for comparison with experiments of high-temperature 
superconductors~\cite{krishna}. At present, the analysis of the thermal 
conductivity has been performed in the real time 
formalism~\cite{ssw,lattice2}, and the extension to a lattice 
analysis is not trivial. We hope to return to this important 
issue in a future publication.

\section*{Acknowledgements}

We wish to thank J. Alexandre and 
A. Campbell--Smith for useful discussions, and also   
the anonymous referee for very constructive comments. 
A.M.
would like to thank M. Hott for providing ref. \cite{reuter}.  
K.F., G.K. and N.E.M. thank S.Hands for useful discussions. 
K.F and G.K also thank G.Tiktopoulos for useful comments 
on several aspects of this work.
The work of N.E.M.  is partially supported by P.P.A.R.C. (U.K.)  under an
advanced fellowship, and that of A.M. by P.P.A.R.C. (U.K.).  K.F. and
G.K. would like to thank the EU for financial support (TMR project,
contract no.: FMRX-CT97-0122).

\appendix

\section{ Calculation of the One-loop  Vacuum Polarization in the 
presence of the External Magnetic Field}

The photon polarization tensor for QED in the presence of the 
electromagnetic field was first performed by Tsai \cite{tsai,reuter}. Here, for
the sake of completeness, we outline his calculations but we will work
in three-dimensions instead of four-dimensions.

Let us begin with the one-loop vacuum polarization graph
(\ref{f2}). The polarization tensor to this approximation is given
by 
\be
\Pi_{\m \n} (p) = -g^2 \int \frac{d^3k}{(2\pi)^3} \tr \left[ \gamma^\m
\tilde{S}(k) \gamma^\n \tilde{S}(p-k) \right]
\label{b.1}
\ee
where the Fermion propagator is the one in the presence of the
constant external magnetic field \cite{schwinger}
\bea
\tilde{S}_F(k) &=& i \int^\infty_0 ds 
e^{- s ( k_0^2 + m^2 +\bbox{k}^2 \frac{\tanh
z}{z})} [ ( m- \g \cdot  k) - i ( \g_1 k_2 - \g_2 k_1) \tanh z ]
( 1- i \g_1 \g_2 \tanh z) \nn \\
&=&  i \int_0^\infty {ds}e^{-s ( m^2 + k_0^2 + {\bbox{ k}}^2
\frac{ \tanh z}{z})} ( ( m - \g_0 k_0) ( 1- i \g_1 \g_2 \tanh z) - (
{\bbox{ \g \cdot k}} ) \frac{1}{ \cosh^2 z})
\label{b.2}
\eea
where $z \equiv eBs$. Accordingly, (\ref{b.2}) leads us to 
\bea
\Pi_{\m \n} (p) = &&-g^2 \int \frac{d^3k}{(2\pi)^3} \int^\infty_0 ds_1 
\int_0^\infty ds_2 e^{- (\chi_0(s_1,k) + \chi_0 (s_2, p-k))} \nn \\
\times && \tr \left[ \gamma^\m (( m - \g_0 k_0) 
( 1- i \g_1 \g_2 \tanh z_1) - (
{\bbox{ \g \cdot k}} ) \frac{1}{ \cosh^2 z_1})  \right. \nn \\
&& \qquad 
 \left. \gamma^\n (( m - \g_0 (p-k)_0) ( 1- i \g_1 \g_2 \tanh z_2) - (
{\bbox{ \g \cdot (p-k)}} ) \frac{1}{ \cosh^2 z_2}) \right]
\label{b.3}
\eea
where $z_i=eBs_i, ~~i=1,2$ and 
\be
\chi_0 (s,k) \equiv s ( m^2 + k_0^2 + {\bbox{ k}}^2 \frac{ \tanh z}{z})
\label{b.4}
\ee
Let us now make the change of variables 
\be
s_1 \equiv \frac{1-v}{2}s, \qquad s_2=\frac{1+v}{2}s,
\label{b.5}
\ee
with $s \in [0, \infty)$ and $ v \in [-1,1]$. Accordingly, we get 
\be
\chi_0(s_1,k) + \chi_0 (s_2, p-k) = s [ \phi_0(p) + \phi_1(p,k)]
 \label{b.6}
\ee
with 
\bea
\phi_0(p) &&\equiv m^2 + \frac{1-v^2}{4} p_0^2 + \frac{\cosh zv - \cosh
z}{2 z \sinh z} {\bbox{p}}^2 \nn \\
\phi_1(p,k) &&\equiv ( k_0 - \frac{1+v}{2} p_0)^2 + \frac{\tanh z_1 + \tanh
z_2}{z} \left( \bbox{k} - \frac{\tanh z_2}{\tanh z_1 + \tanh
z_2} \bbox{p} \right)^2
\label{b.6.5}
\eea
The loop integrals can then be performed very easily via standard
Gaussian integrations some of which are listed below:
\bea
I_0 \equiv &&\int \frac{d^3k}{(2\pi)^3} e^{-s \phi_1(p,k)} = 
\frac{1}{(4\pi s)^{\frac{3}{2}}}{\frac{z}{\sinh z}} \cosh z_1 \cosh
z_2 
\label{b.6.6} \\
\int \frac{d^3k}{(2\pi)^3}&& e^{-s \phi_1(p,k)}k_0 =\frac{1+v}{2} p_0
I_0 \label{b.6.7}\\
\int \frac{d^3k}{(2\pi)^3}&& e^{-s \phi_1(p,k)}{\bbox {k}} = \frac{\tanh
z_2}{\tanh z_1 + \tanh z_2} {\bbox{ p}} I_0 \label{b.6.8} \\
\int \frac{d^3k}{(2\pi)^3} &&e^{-s \phi_1(p,k)} k_0 {\bbox{ k}} = 
\frac{1+v}{2} \frac{\tanh z_2}{\tanh z_1 + \tanh z_2} I_0 p_0 {\bbox{
p}} \label{b.6.9} \\
\int \frac{d^3k}{(2\pi)^3} &&e^{-s \phi_1(p,k)} k_i k_j = \left[
\left(\frac{ \tanh z_1}{\tanh z_1 + \tanh z_2} \right)^2 p_i p_j -
\frac{z}{s( \tanh z_1 + \tanh z_2)} \delta _{ij} \right]
\label{b.6.10}
\eea
  
We will also need the following identity for the gamma matrices:
\bea
\tr [ ( 1- i \g_1 \g_2 \tanh z) \g_\m \g_\n] &&=- 4( \delta_{\m \n} -
\tanh z ( \delta^\m_1 \delta^\n_2 - \delta^\n_1 \delta^\m_2)) \nn \\
&& = -4( \delta_{\m \n} -
\tanh z \frac{F_{\m \n}}{B})
\label{b.7}
\eea          
where $F_{\m,\n}$ is the covariant representation of the external
magnetic field strength. The other traces can also be evaluated by 
the use of the Dirac algebra (\ref{identities}).

Putting everything together \cite{tsai,reuter} one gets ( after an
integration by parts )
\be
\Pi_{\m \n} (p) = \frac{g^2}{\sqrt{2\pi}} \int \frac{ds}{\sqrt{s} }
\frac{dv}{2} \frac{z}{\sinh z} e^{-s \phi_0} I_{\m \n}
\label{b.8}
\ee
where 
\be
I_{\m \n} = [ (\delta_{\m \n} p^2 - p_\m p_\n) R_0(p) + (
\delta^\perp_{\m \n} p^2_\perp - p_{\perp \m} p_{\perp \n} ) R_1(p)] 
\label{b.9}
\ee
with 
\bea
R_0(p) = ( \cosh zv - v \coth z \sinh zv ) \label{b.9.1} \\
R_1(p) = \frac{2}{\sinh^2 z} [ \cosh z - \cosh zv] - R_0(p)
\label{b.9.2}
\eea

Note that, unlike its four-dimensional counterpart, the vacuum
polarization tensor in three dimensions is not divergent and there is
no need add any counterterms. One can see this by checking the absence
of poles at $s \ra 0,$ which is the place where poles usually show up   
in proper time methods.

\begin{figure}[htb]
\begin{center} 
\begin{picture}(100,60)(0,0) 
\Photon(0,30)(30,30){3}{2.5}
\CArc(50,30)(20,0,360)
\Text(50,50)[]{$\times$}
\Text(50,10)[]{$\times$}
\Photon(70,30)(100,30){3}{2.5}
\end{picture}
\caption{{ One-loop vacuum polarization 
for photons (wavy lines) in $QED_3$. The 
solid lines with crosses represent fermions in the presence of 
an external  magnetic field.}} 
\label{f1}
\end{center}
\end{figure}
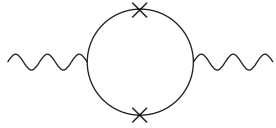

\begin{figure}[hbt]
\begin{center}
\begin{picture}(300,50)(0,0)
\Text(5,30)[]{$\Big($}
\Line(10,30)(70,30)
\Text(22,30)[]{$\times$}
\Text(58,30)[]{$\times$}
\Text(80,31)[]{$\Big)^{-\! 1}$}
\GOval(40,30)(6,8)(0){0}
\CBox(32,23)(48,29){White}{White}
\Text(95,30)[]{$=$}
\Text(110,30)[]{$\Big($}
\Line(115,30)(175,30)
\Text(145,30)[]{$\times$}
\Text(185,31)[]{$\Big)^{-\! 1}$}
\Text(200,30)[]{$-$}
\Line(215,30)(295,30)
\GOval(255,30)(6,8)(0){0}
\CBox(247,23)(263,29){White}{White}
\PhotonArc(255,30)(25,9,79){-3}{2.5}
\PhotonArc(255.4,30)(25,101,180){-3}{2.5}
\Vertex(255,55){5}
\GCirc(230,30){1}{0.2}
\GCirc(280,30){4}{0.2}
\Text(240,30)[]{$\times$}
\Text(270,30)[]{$\times$}
\end{picture}
\caption{The Schwinger-Dyson equation for the fermion self-energy.
The curly line indicates the $U_S(1)$ statistical photon. 
Solid lines with crosses represent fermions in the presence of the external 
magnetic field. Blobs indicate quantum corrections (loops), 
which are ignored 
in the ladder approximation. 
Quantum dynamics of the 
electromagnetic field has been suppressed.\label{f2}} 
\end{center}
\end{figure}

\newpage

\begin{figure}
\centerline{\hbox{\psfig{figure=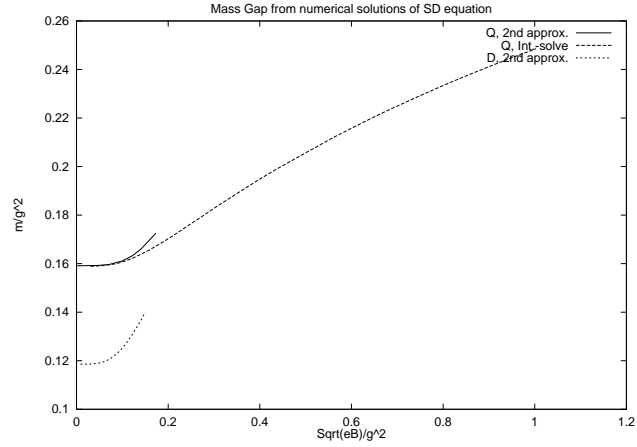,height=6cm,angle=-90}}}
\caption[f3]{Solution of Schwinger-Dyson equations for the quenched
and dynamical fermions.} 
\label{f3}
\end{figure}

\begin{figure}
\centerline{\hbox{\psfig{figure=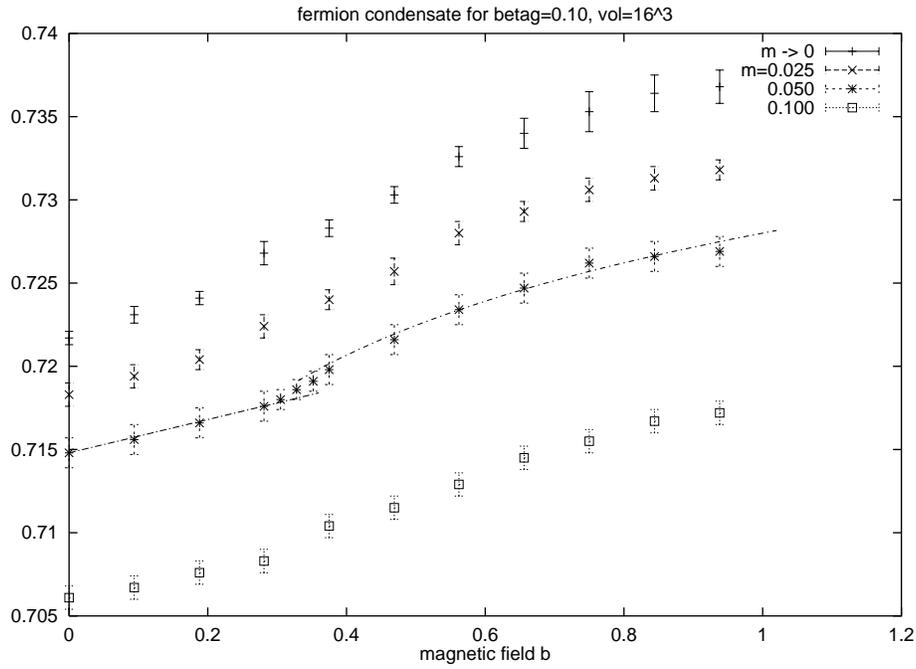,height=9cm,angle=-90}}}
\caption[f4]{$<{\overline \Psi} \Psi>$ versus the magnetic field 
strength at strong coupling for three masses and extrapolation to
the zero mass limit.}
\label{f4}
\end{figure}
\begin{figure}
\centerline{\hbox{\psfig{figure=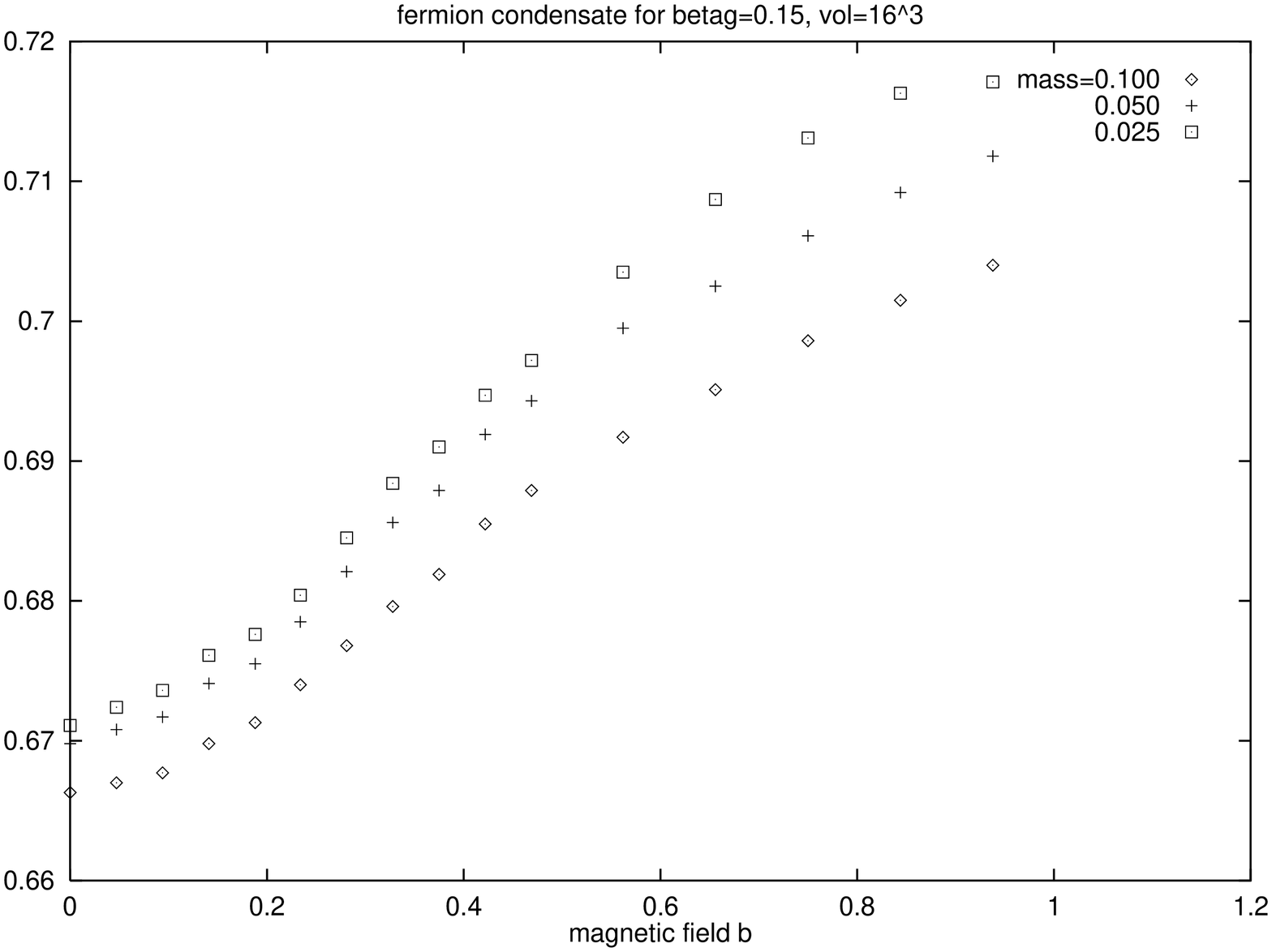,height=9cm,angle=-90}}}
\caption[f5]{$<{\overline \Psi} \Psi>$ versus magnetic field  strength 
at intermediate coupling for three masses.}
\label{f5}
\end{figure}
\begin{figure}
\centerline{\hbox{\psfig{figure=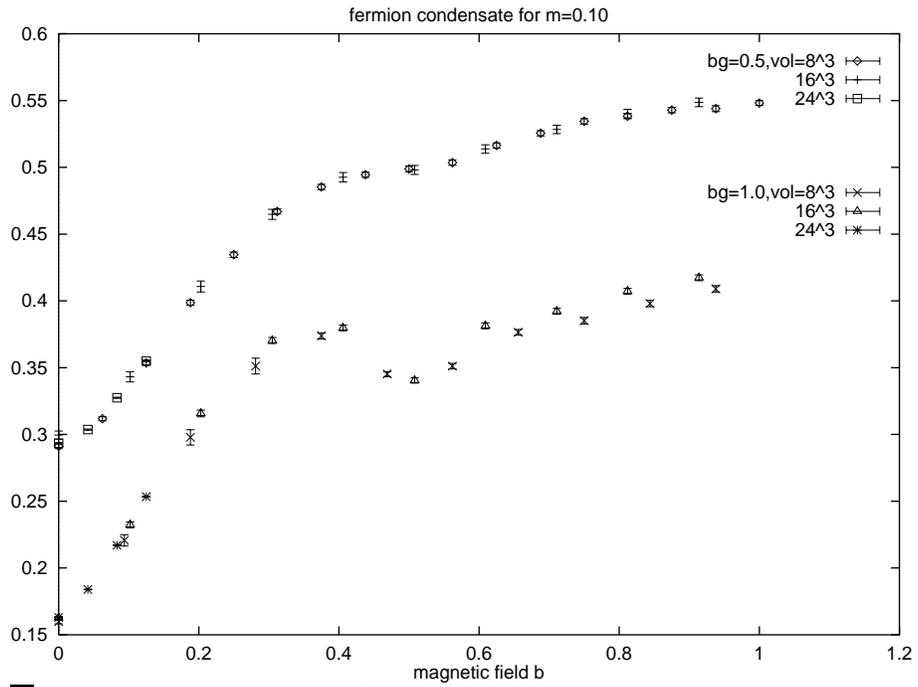,height=9cm,angle=-90}}}
\caption[f6]{$<{\overline \Psi} \Psi>$ versus the magnetic field 
for two small values of the gauge coupling constant and three volumes.}
\label{f6}
\end{figure}
\begin{figure}
\centerline{\hbox{\psfig{figure=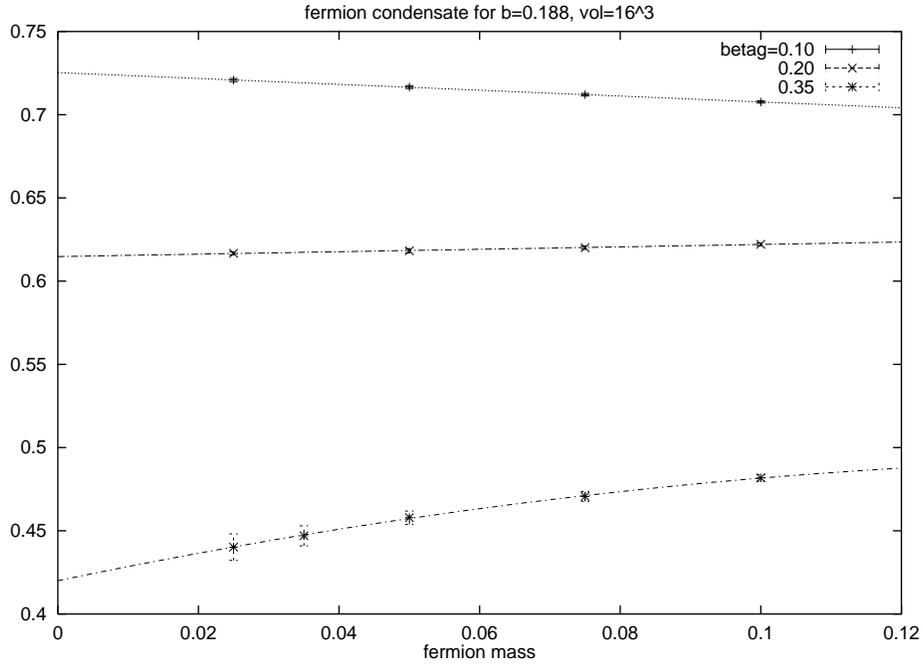,height=9cm,angle=-90}}}
\caption[f7]{$<{\overline \Psi} \Psi>$ versus m for a typical
value of the magnetic field strength and various values of 
$\beta_G.$} 
\label{f7}
\end{figure}
\begin{figure}
\centerline{\hbox{\psfig{figure=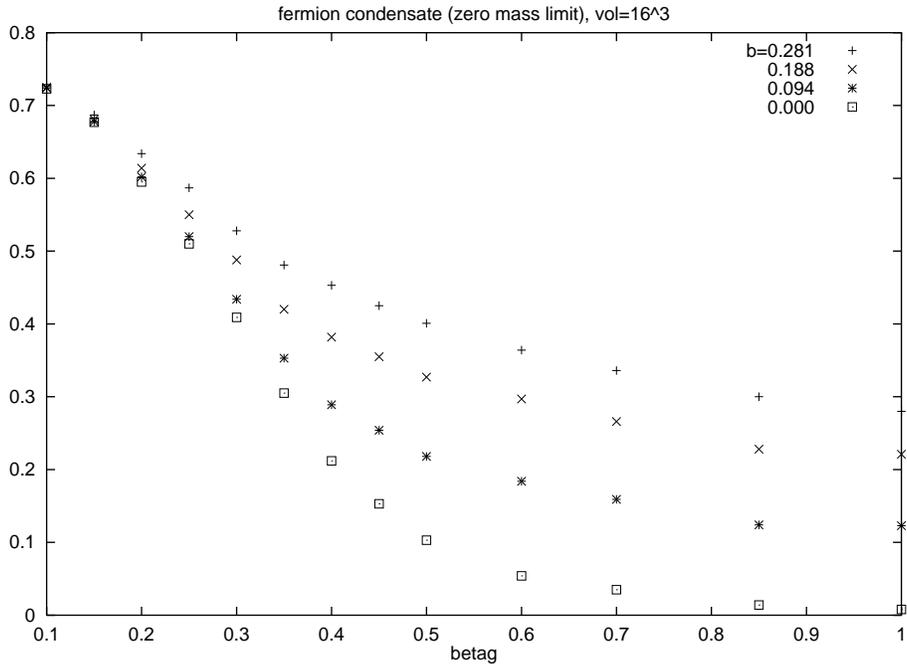,height=9cm,angle=-90}}}
\caption[f8]{$<{\overline \Psi} \Psi>$ versus $\beta_G$ n the
zero mass limit for four values of the magnetic field strength.}
\label{f8}
\end{figure}

\begin{figure}
\centerline{\hbox{\psfig{figure=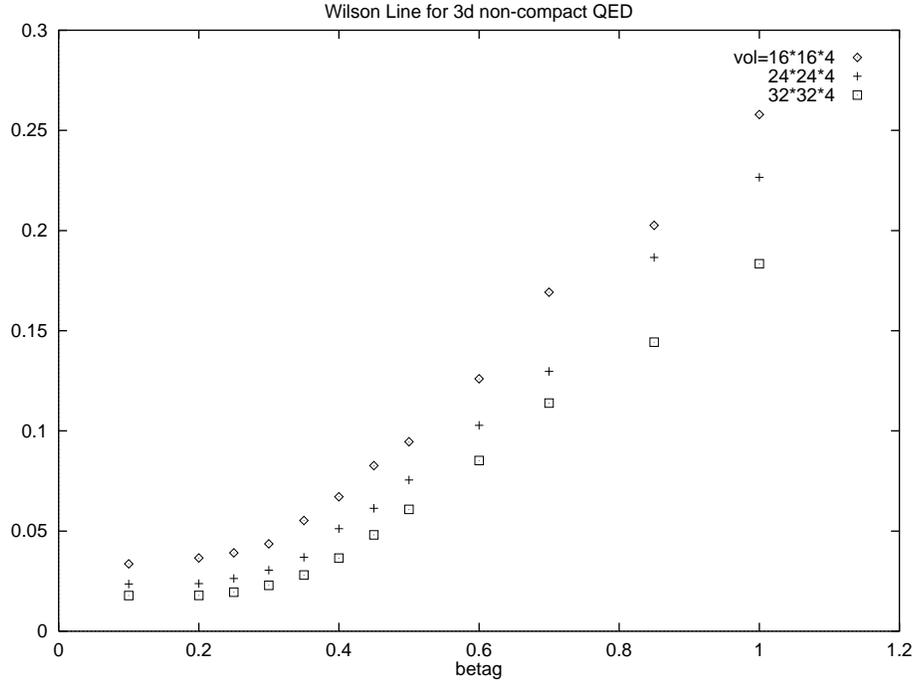,height=9cm,angle=-90}}}
\caption[f9]{Wilson line versus $\beta_G$ for $N_\tau=4$ and
three spatial sizes.} 
\label{f9}
\end{figure}

\begin{figure}
\centerline{\hbox{\psfig{figure=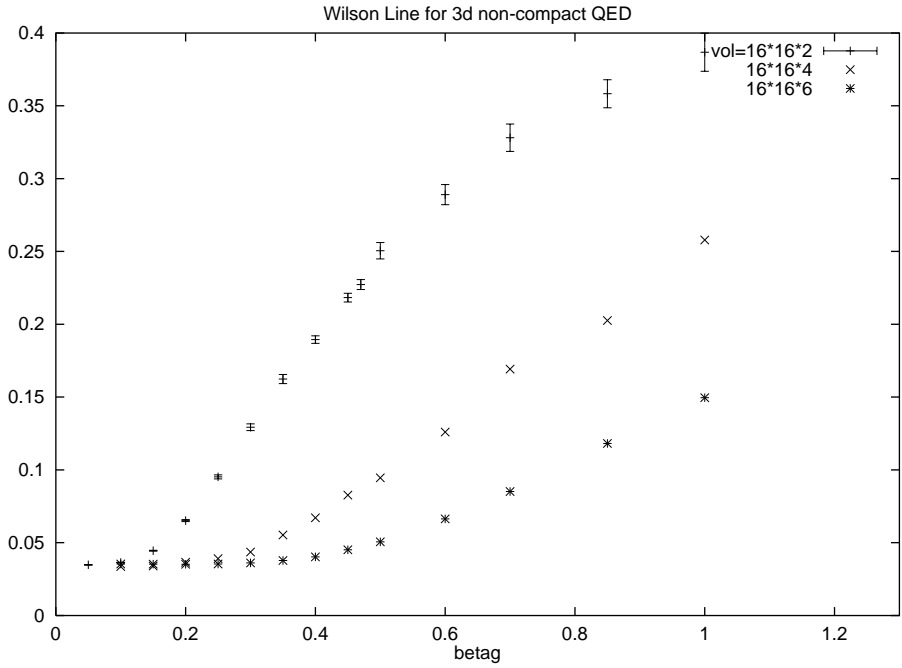,height=9cm,angle=-90}}}
\caption[f10]{Wilson line versus $\beta_G$ for $N_\tau=2$, 
$N_\tau=4$ and $N_\tau=6.$}
\label{f10}
\end{figure}
\begin{figure}
\centerline{\hbox{\psfig{figure=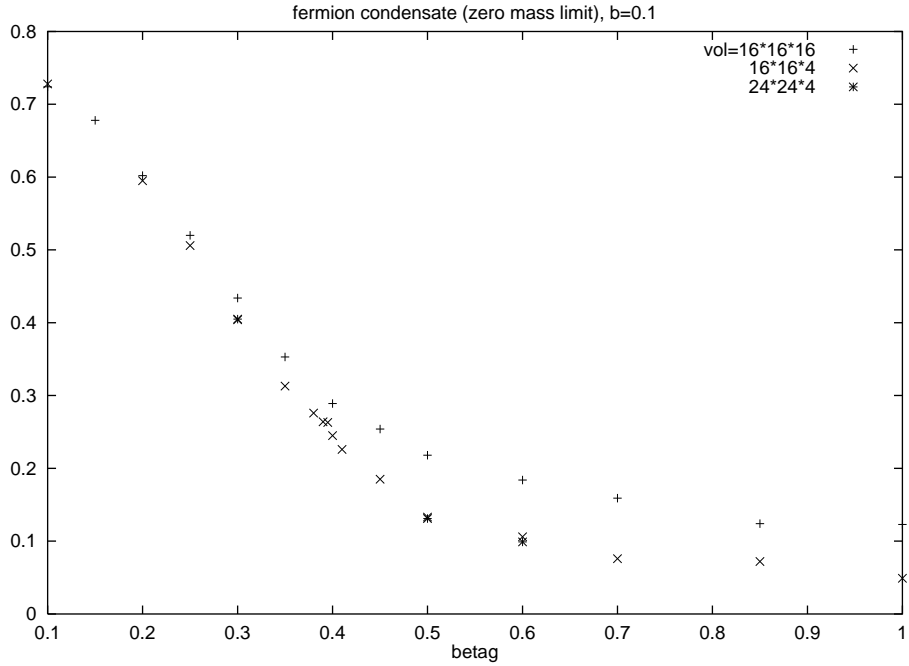,height=9cm,angle=-90}}}
\caption[f11]{Condensate versus $\beta_G$ for b=0.1.
Comparison of zero temperature with finite temperature.}
\label{f11}
\end{figure}
\begin{figure}
\centerline{\hbox{\psfig{figure=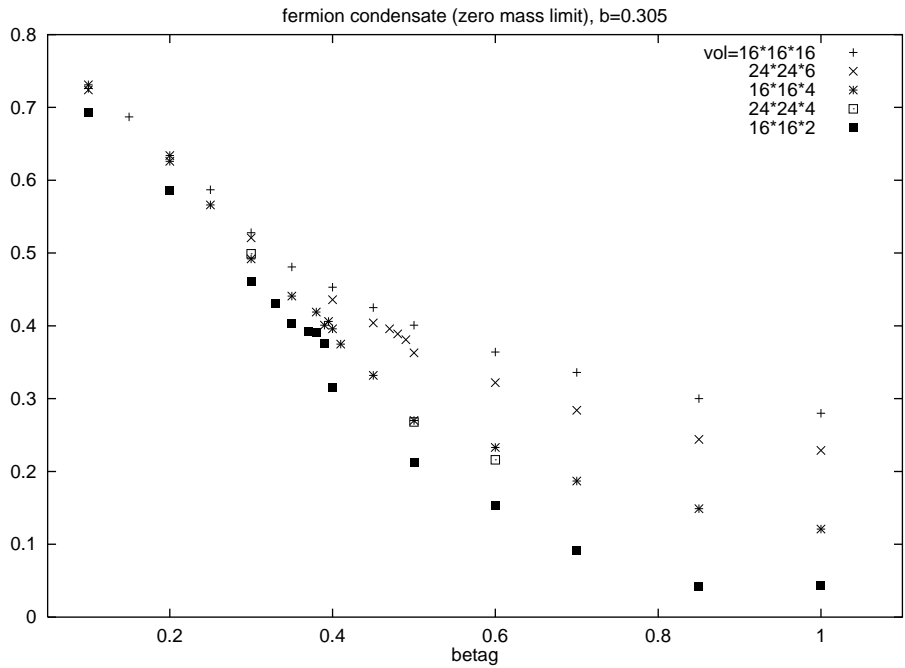,height=9cm,angle=-90}}}
\caption[f12]{Condensate versus $\beta_G$ for b=0.3.
Comparison of zero temperature with finite temperature. 
The error bars(not shown) are almost of the size of the symbols.}          
\label{f12}
\end{figure}

\begin{figure}
\centerline{\hbox{\psfig{figure=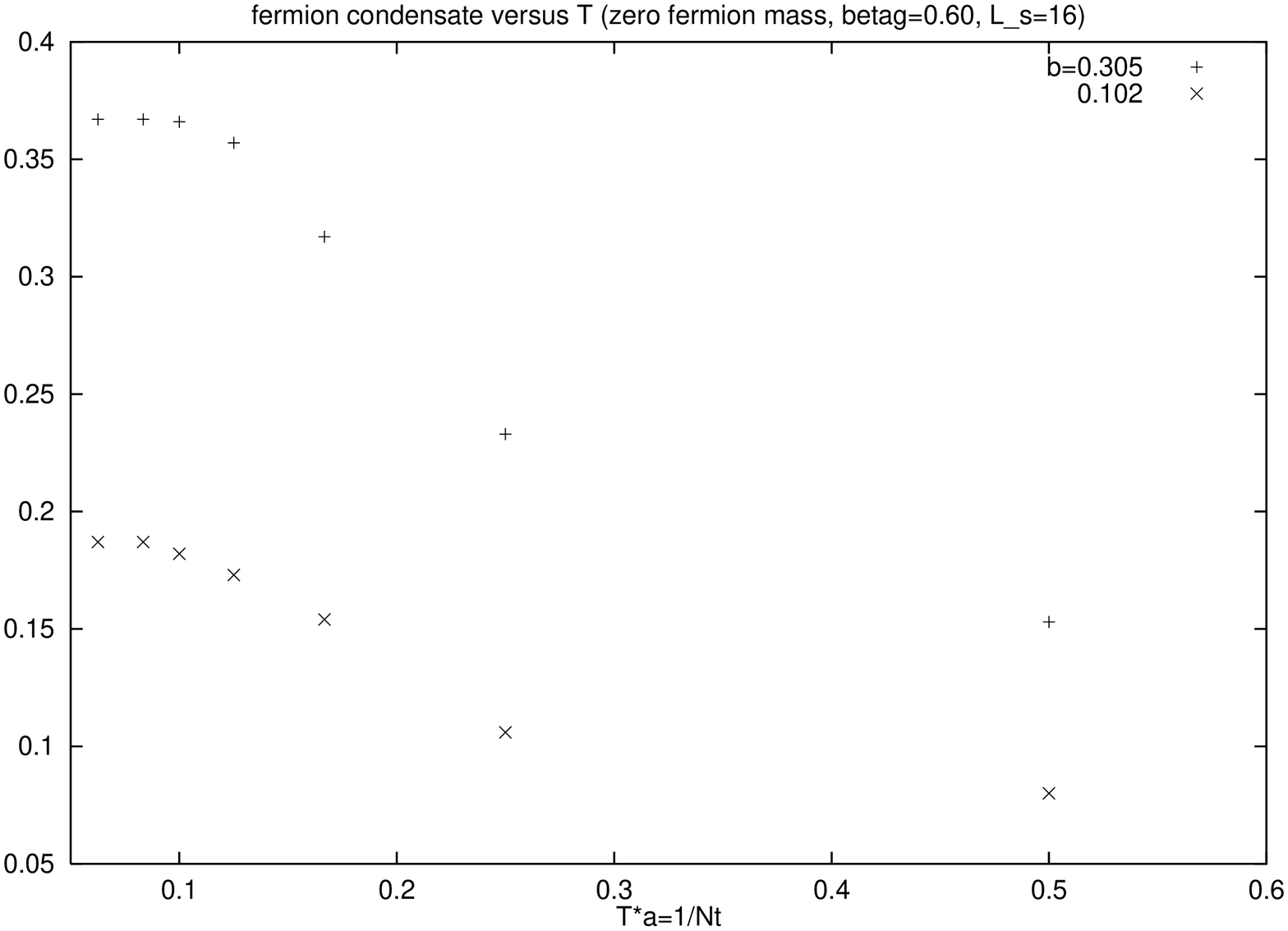,height=9cm,angle=-90}}}
\caption[f13]{$<{\overline \Psi} \Psi>$ versus the temperature 
for two values of the magnetic field.}
\label{f13}
\end{figure}
\begin{figure}
\centerline{\hbox{\psfig{figure=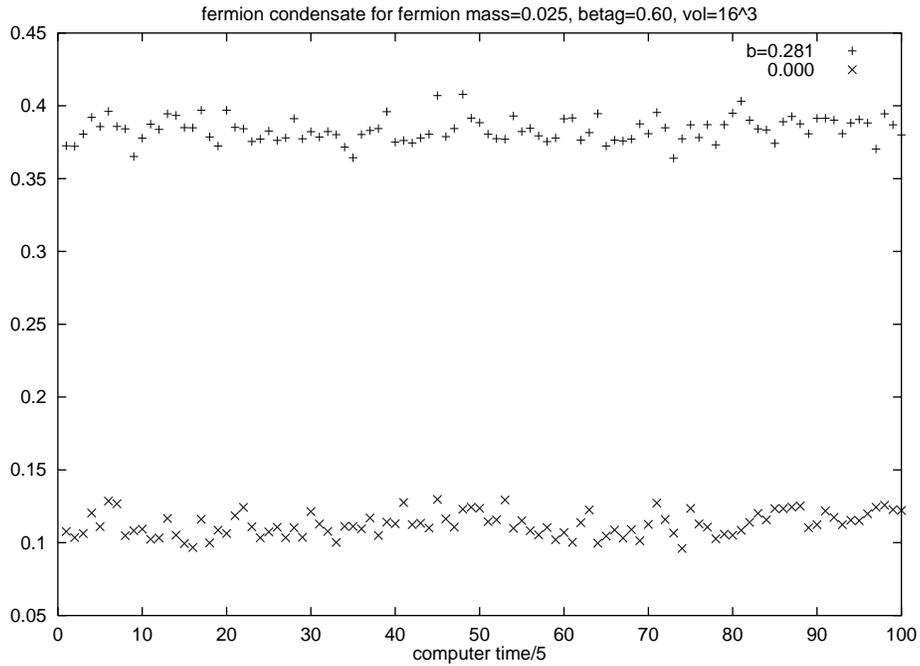,height=9cm,angle=-90}}}
\caption[f14]{Time evolution of $<{\overline \Psi} \Psi>$ 
for a symmetric $(16^3)$ lattice and two values of the magnetic field.} 
\label{f14}
\end{figure}
\begin{figure}
\centerline{\hbox{\psfig{figure=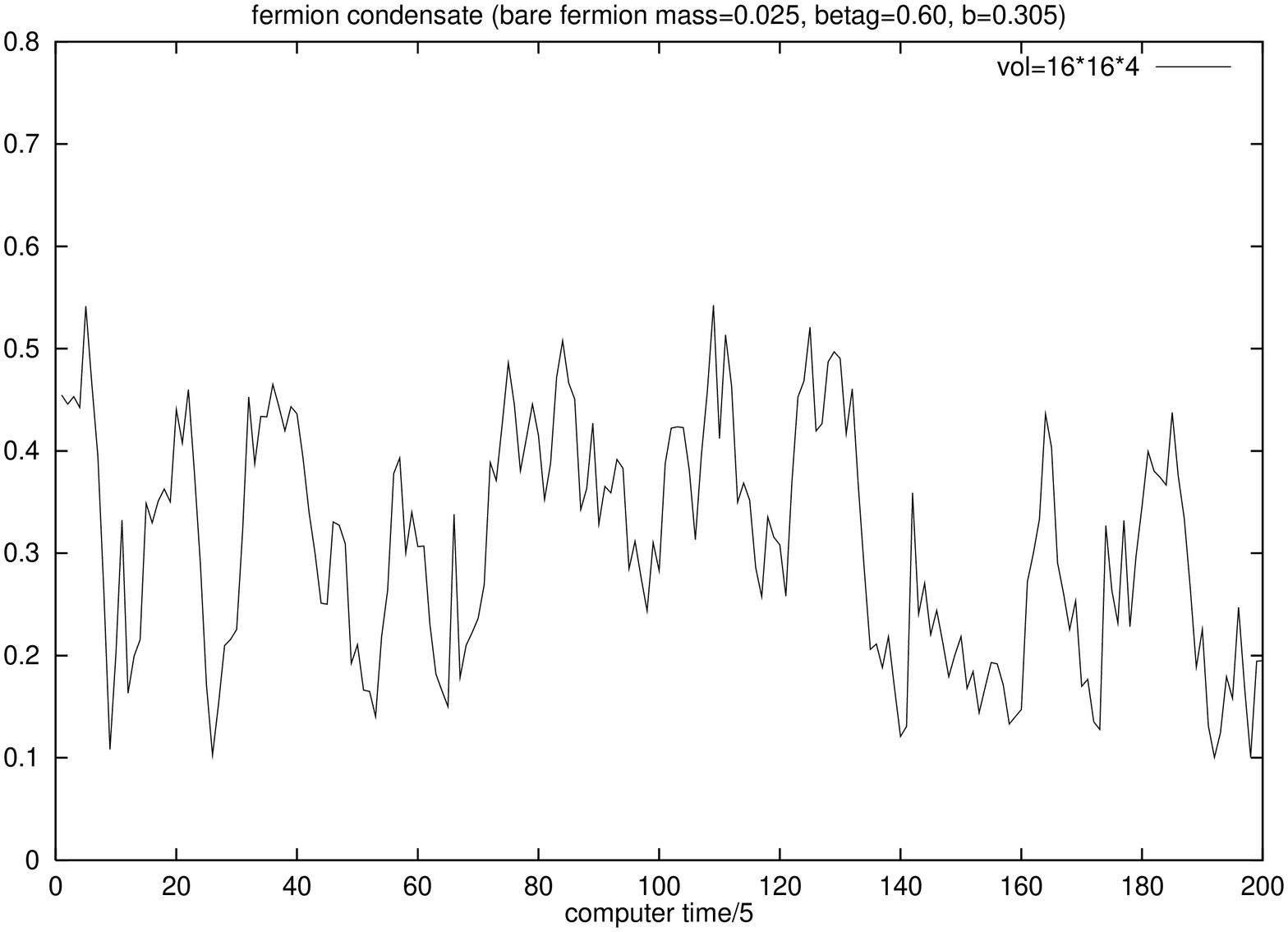,height=9cm,angle=-90}}}
\caption[f15]{Time evolution of $<{\overline \Psi} \Psi>$  
for a $16^2*4$ lattice.} 
\label{f15}
\end{figure}
\begin{figure}
\centerline{\hbox{\psfig{figure=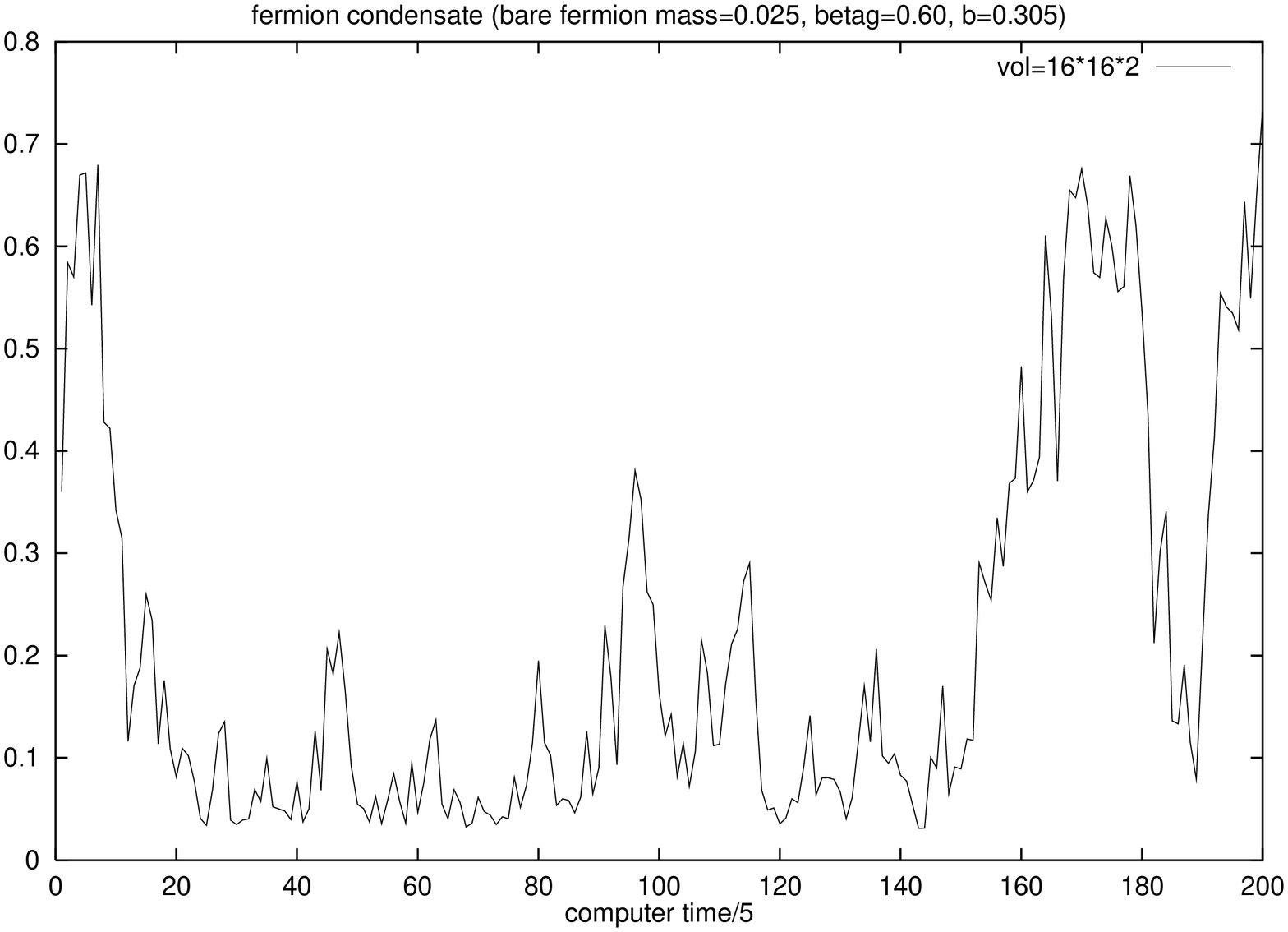,height=9cm,angle=-90}}}
\caption[f16]{Time evolution of $<{\overline \Psi} \Psi>$  
for a $16^2*2$ lattice.} 
\label{f16}
\end{figure}
\begin{figure}
\centerline{\hbox{\psfig{figure=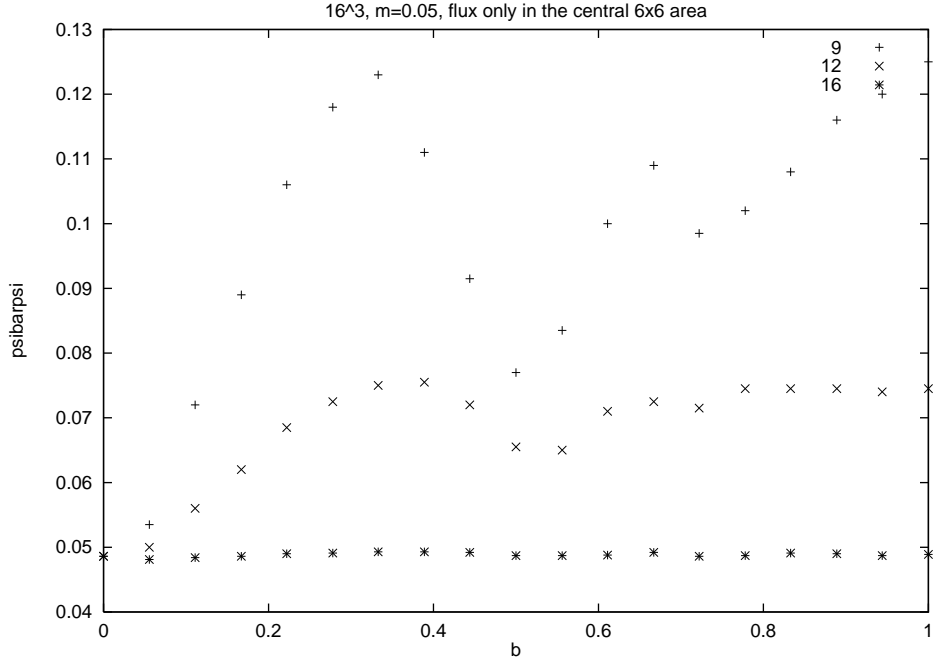,height=9cm,angle=-90}}}
\caption[f17]{$<{\overline \Psi} \Psi>$ versus magnetic field strength  
where the flux is non zero only in a central region extending over 
6x6 plaquettes. The condensate at sites labeled 9, 12, 16 (see text) is shown. 
The corresponding distances from the center of the flux tube are 0, 3, 7.}
\label{f17}
\end{figure}
\begin{figure}
\centerline{\hbox{\psfig{figure=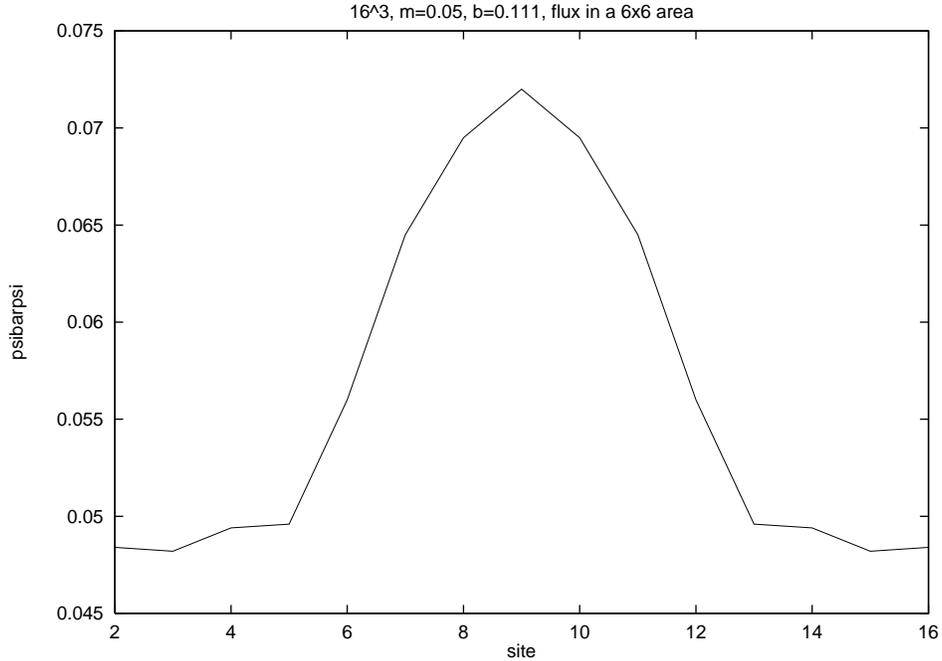,height=9cm,angle=-90}}}
\caption[f18]{$<{\overline \Psi} \Psi>$ along a straight line
passing from the center of the lattice if the magnetic field 
parameter $b$ is set to 0.111. The central region of 
non-zero flux is 6x6.} 
\label{f18}
\end{figure}

\end{document}